\documentclass[a4paper,fleqn,usenatbib]{mnras}

\usepackage{newtxtext,newtxmath}

\usepackage[T1]{fontenc}
\usepackage{ae,aecompl}

\usepackage{graphicx}	
\usepackage{amsmath}	
\usepackage{amssymb}	
\usepackage{subfig}
\usepackage{multirow}
\usepackage{caption}
\usepackage{booktabs}

\title[Mini-BALs to BALs]{HST and Ground-Based Spectroscopy of Quasar Outflows: From Mini-BALs to BALs}

\author[E. A. Moravec et al.]{
E. A. Moravec,$^{1}$\thanks{E-mail: emoravec@ufl.edu} 
F. Hamann,$^{2}$ D. M. Capellupo,$^{3}$ S. M. McGraw,$^{4}$ J. C. Shields,$^{5}$
\newauthor{ P. Rodr\'iguez Hidalgo$^{6,7}$}
\\
$^{1}$Department of Astronomy, University of Florida, Gainesville, FL 32607, USA\\
$^{2}$Department of Physics \& Astronomy, University of California - Riverside, CA 92521, USA\\
$^{3}$Department of Physics, McGill University, Montr{\'e}al, QC H3A 0G4, Canada\\
$^{4}$Department of Astronomy \& Astrophysics, Pennsylvania State University, University Park, PA 16802, USA\\
$^{5}$Department of Physics \& Astronomy, Ohio University, Athens, OH 45701, USA\\
$^{6}$Department of Physics \& Astronomy, York University, Toronto, ON, M3J 1P3, Canada\\
$^{7}$Department of Physics \& Astronomy, Humboldt State University, Arcata, CA 95521, USA\\
}

\date{Accepted 2017 March 27; Received 2017 February 23; in original form 2017 January 20} 

\pubyear{2017}

\begin{document}
\label{firstpage}
\pagerange{\pageref{firstpage}--\pageref{lastpage}}
\maketitle

\begin{abstract}
Quasar outflows have been posited as a mechanism to couple super-massive black holes to evolution in their host galaxies. We use multi-epoch spectra from the Hubble Space Telescope and ground-based observatories to study the outflows in seven quasars that have CIV outflow lines ranging from a classic BAL to weaker/narrower ``mini-BALs''  across rest wavelengths of at least 850 $\AA$ to 1650 $\AA$. The CIV outflow lines all varied within a time frame of $\leq$ 1.9 yrs (rest). This includes equal occurrences of strengthening and weakening plus the emergence of a new BAL system at $-$38,800 km/s accompanied by dramatic strengthening in a mini-BAL at $-$22,800 km/s. We infer from $\sim$1:1 doublet ratios in PV and other lines that the BAL system is highly saturated with line-of-sight covering fractions ranging from 0.27 to 0.80 in the highest to lowest column density regions, respectively. Three of the mini-BALs also provide evidence for saturation and partial covering based on $\sim$1:1 doublet ratios. We speculate that the BALs and mini-BALs form in similar clumpy/filamentary outflows, with mini-BALs identifying smaller or fewer clumps along our lines of sight. If we attribute the line variabilities to clumps crossing our lines of sight at roughly Keplerian speeds, then a typical variability time in our study, $\sim$1.1 yrs, corresponds to a distance $\sim$2 pc from the central black hole. Combining this with the speed and {\it minimum} total column density inferred from the PV BAL, $N_H\ga 2.5\times 10^{22}$ cm$^{-2}$, suggests that the BAL outflow kinetic energy is in the range believed to be sufficient for feedback to galaxy evolution. 
\end{abstract}

\begin{keywords}
galaxies: active - quasars: general - quasars: absorption lines
\end{keywords}



\section{Introduction}\label{intro}
Observations indicate that supermassive black holes (SMBHs) and their host galaxies evolve together, regulating one another's growth as seen through strong correlations between the SMBH mass and other physical properties of the host galaxy (see \citealt{Kormendy13} for a recent review, \citealt{King15}). Quasar outflows have been posited as a mechanism to physically couple the central SMBHs to their host galaxies through kinetic energy `feedback,' which can clear the host galaxy of interstellar gas, terminate the star formation, and halt the infall of matter into both the galaxy and the central SMBH (\citealt{Kauffmann00}, \citealt{Granato04}, \citealt{DiMatteo05}, \citealt{Hopkins10}, \citealt{Fabian12} for a recent review). The outflow kinetic energies required for feedback to contribute to galaxy evolution in this way, expressed as a luminosity relative to the quasar bolometric luminosity $L_K/L$, are expected to be $\ga$0.005 (\citealt{Hopkins10}) to $\ga$0.05 (\citealt{Scannapieco04}, \citealt{DiMatteo05}, \citealt{Prochaska09}). It is not yet known if the accretion disk outflows from quasars have kinetic energies sufficient for feedback. Thus, better observational constraints are still needed on basic properties such as outflow locations, spatial structure, and total column densities.

Accretion disk outflows are commonly observed via blueshifted broad absorption lines (BALs) in the rest frame UV. BALs are generally defined to have widths of $\gtrsim$ 2000 km/s (\citealt{Weymann91}, \citealt{Reichard03}, \citealt{Trump06}, \citealt{Capellupo13}). However, quasar outflows also produce a wide variety of narrow absorption lines (NALs) with widths of less than a few hundred km/s and so-called mini-BALs with intermediate widths between BALs and NALs (\citealt{Hamann97}, \citealt{Hamann04}, \citealt{Vestergaard03}, \citealt{Paola09}, \citealt{Paola11}, \citealt{Hamann11}, \citealt{Misawa07}, \citealt{Misawa14}, Rodr\'{i}guez Hidalgo et al. in prep). The outflow NALs and mini-BALs are together more common than the well-studied BALs in quasar spectra. Specifically, $\sim$50\% of bright SDSS quasars contain either a CIV NAL or mini-BAL outflow whereas $\sim$20\% contain a CIV BAL outflow (\citealt{Hamann12}, \citealt{Paola09}). Of the known outflow types, mini-BALs are the least studied and least understood. Further, We still do not have a clear understanding of how BALs, NALs, and mini-BALs fit together into a unified picture of the quasar outflow phenomenon. 

One limitation is that important outflow lines, such as OVI 1031,1038 $\AA$ appear at wavelengths less than 1216 $\AA$  where they are susceptible to contamination by the rich forest of Ly$\alpha$ absorption lines below $\sim$1216 $\AA$ . The problems with Ly$\alpha$ forest contamination are more severe for weak mini-BALs than for BALs. However, measurements at these wavelengths are important in order to obtain the absorber ionizations and column densities. Far-UV observations of BAL quasars have shown that outflow ionizations are generally quite high, characterized by stronger absorption in OVI 1031,1038 $\AA$ compared to CIV (\citealt{Hamann98}, \citealt{Arav01}, \citealt{Leighly09}, \citealt{Baskin13}, Herbst et al., in prep.). These observations have also shown that a variety of other lines, such as CIII 977 $\AA$ , SVI 933,944 $\AA$, SIV 1063,1073 $\AA$, and PV 1118,1128 $\AA$, can be available for additional ionization and column density constraints (see refs. above, also \citealt{Dunn12}, \citealt{Capellupo13}, \citealt{Capellupo14} and in prep.). 

\begin{table*}
	\caption{Quasar Properties: The basic properties of our sample of quasars  -- the full quasar name, redshift, Balnicity Index (BI), FWHM of the outflow feature for which the quasar was chosen, and the bolometric luminosity.}
	\centering
	\label{table:quasar_properties}
	\begin{tabular}{*{6}{c}} 
	\hline
	Quasar & $z_{em}$ & $z_{abs}$ & BI & FWHM & $L$ \\
	 & & (km/s) & (km/s) & (erg/s) \\
	\hline
	J020845.54+002236.07 &1.8983 & 1.6784 & 0 & 1551 & 3.9E47 \\
	J090552.41+025931.46 & 1.8268 & 1.6208 & 40 & 2518 & 3.2E47\\
	J090924.01+000211.04 & 1.8781 & 1.6707 & 277 & 2003 & 3.5E47\\
	J100023.49+124705.43 &1.6866 & 1.5080 & 56 & 1777 & 1.5E47\\
	J100128.61+502756.90 &1.8410 & 1.7682 & 147 & 2709 & 2.2E47\\
	J103112.24+380717.23 &1.9032 & 1.6846 & 0 & 986 & 3.7E47\\
	J130136.13+000157.86 &1.7892 & 1.6915 & 4338 & 7593 & 1.8E47\\
	\hline
	\end{tabular}
\end{table*}

\begin{table}
	\centering
	\caption{Telescope and Instrument Summary}
	\label{table:instrum}
	\begin{tabular}{*{6}{c}} 
	\hline 
	Telescope & Spectrograph & $\lambda$ Range ($\AA$ )& Resolution \\
	\hline
	HST 2.4-m & COS & 2450 - 3175 & $\sim$2000 \\ 
	MDM 2.4-m & CCDS & 3200 - 4800 & $\sim$1200\\
	KPNO 2.1-m & GoldCam & 3700 - 6100 & 1300 - 1500 \\
	SDSS 2.5-m & SDSS & 3800 - 9200 & 1800 - 2200 \\
	BOSS 2.5-m & BOSS & 3650 - 10400  & 1500 - 2500\\
	\hline
	\end{tabular}
\end{table}

The low-abundance PV lines are particularly valuable as indicators of the true total column densities in situations where the commonly measured high-abundance lines like CIV and OVI are saturated (\citealt{Hamann98}, \citealt{Hamann03}, \citealt{Capellupo14}, McGraw et al., in prep., Herbst et al., in prep.). Photoionization models that assume solar abundance ratios and standard ionizing quasar spectra indicate that the CIV optical depths are at least 100 times and up to $\sim$1200 times larger than PV (\citealt{Hamann98}, \citealt{Leighly09}, \citeyear{Leighly11}, \citealt{Borguet12}). Thus, the mere presence of PV absorption in the spectrum implies that CIV is extremely saturated and, more importantly, that the total outflow column densities are 2-3 dex larger than would be inferred from CIV or OVI alone. The minimum total column densities inferred for BAL outflows using this PV analysis have been estimated to be $\log (N_H) \gtrsim 22.0$ cm$^{-2}$ and $\gtrsim$ 22.3 cm$^{-2}$ (\citealt{Hamann98}, \citealt{Capellupo14} respectively).

Studies of BAL and NAL outflows have also shown that important constraints on the structure and locations of the flows can be obtained from line variability. This information is essential in order to determine the outflow kinetic energies and their viability for feedback to galaxy evolution (e.g., \citealt{Hamann11}, \citealt{Leighly11}, \citealt{Capellupo14}). However, a major uncertainty in this analysis is the unknown physical cause of the line variations. Two commonly discussed causes are 1) the motions of absorbing gas across our lines of sight to the quasar light source and 2) the change in the outflow ionization state caused by changes in the ionizing flux emitted by the quasar (\citealt{Barlow93}, \citealt{Gabel05}, \citealt{Misawa07c}, \citealt{Hamann08}, \citealt{Gibson08}, \citealt{Leighly09}, \citealt{Moe09}, \citealt{Hamann11}, \citealt{Hall11}, \citealt{Arav12}, \citealt{Vivek12}, \citealt{Capellupo12}, \citealt{Trevese13}, \citealt{Misawa14}, \citealt{Capellupo14}, \citealt{Vivek14}, \citealt{Arav15}, \citealt{Grier15}, \citealt{Rogerson16}). If there is evidence of a changing covering fraction through varying optically thick lines, a natural explanation is crossing clouds and the cloud's movement across the source will dictate the depth of the saturated line. Analysis of line ratios can reveal whether or not the lines are optically thick either through doublet line ratios or comparing the strengths of lines from abundant elements (like CIV) to that of those expected to be weaker such as PV and SiIV (\citealt{Hamann08}, \citealt{Capellupo12}, \citealt{Capellupo14}, McGraw et al., in prep.). On the other hand, line variations caused by changes in the ionization often cannot be ruled out. In some cases, different variability behaviours between lines of different ions might signal ionization changes, and/or coordinated variabilities between distinct outflow velocities components (in the same ion/transition) are perhaps best explained by global outflow ionization changes caused by changes in the ionizing flux.

In the situation where the inferred cause of variability is moving clouds from the detection of PV, the crossing speeds of the clouds can be assumed to be roughly Keplerian (tied to the depth of the gravitational potential well near the central black hole). \citealt{Capellupo14} found a crossing time that then yielded an upper limit on the outflow radial distances of $\lesssim$ 3.5 pc from the black holes and total outflow kinetic energies of $\sim$4x10$^{54}$ ergs. This energy derived for a BAL outflow might be sufficient to drive feedback and affect the host galaxies. 

In this paper, we present Hubble Space Telescope (HST) and ground-based spectra of 7 quasars at redshifts $z_{em}\sim 1.8$ that have a range in outflow line strengths from weak mini-BALs to a strong BAL. We also obtained 5 to 9 ground-based observations per quasar to examine variability in the CIV outflow lines. The wavelengths covered by our study are from at least 850 $\AA$ to 1650 $\AA$ in the quasar rest frame. This is the first study to target specifically mini-BALs and weak BALs in the far-UV for comparisons to the stronger and more often measured BALs. These data provide improved constraints on the outflow locations and ionizations for a range of outflow types and the total column density, mass loss rate, and kinetic energy for the strong BAL.  

In Sections \ref{sample} and \ref{data} below, we describe the sample selection, data sets, and new observations. Section \ref{results} presents the results and analysis based on absorption line fitting and variability measurements. Section \ref{disc} discusses the broader implications for the global picture of quasar outflows, specifically, the structure of quasars outflows, the relationship of mini-BALs to BALs, and the possible role of these outflows for feedback to galaxy evolution. Section \ref{summary} provides a brief summary. 

\section{Quasar Sample} \label{sample}
\begin{figure*}
\centering
	\begin{minipage}{\linewidth}
 		\includegraphics[width=\linewidth, height= 0.35\linewidth, keepaspectratio]{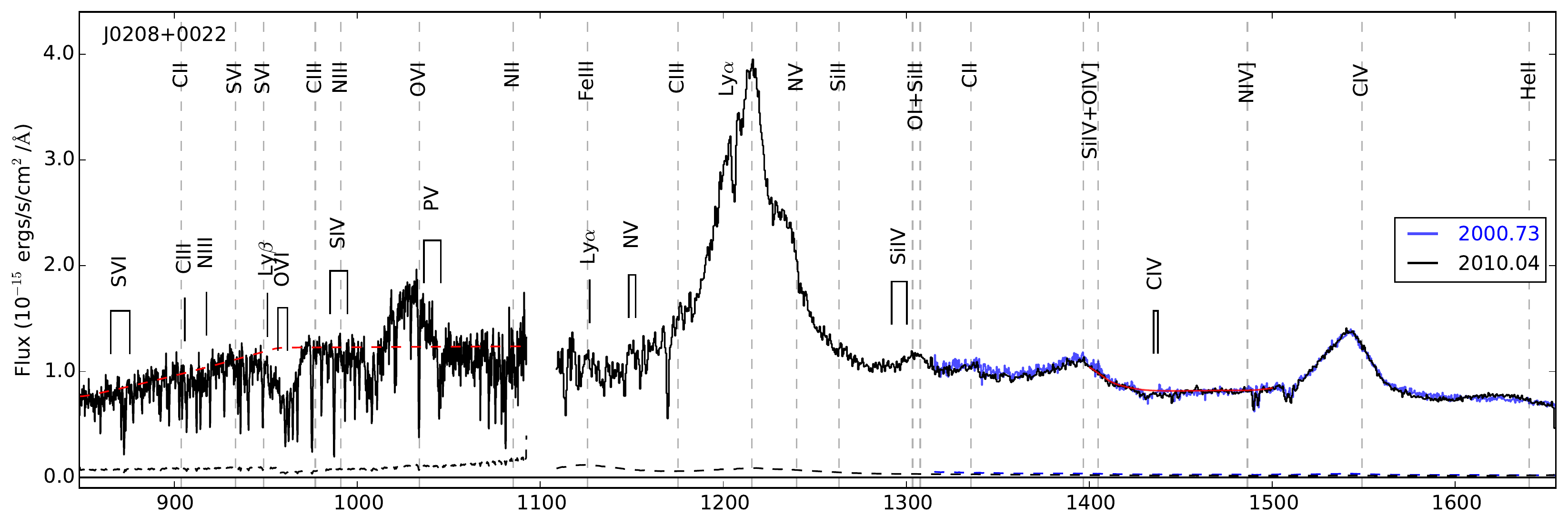}
 		\includegraphics[width=\linewidth, height= 0.35\linewidth, keepaspectratio]{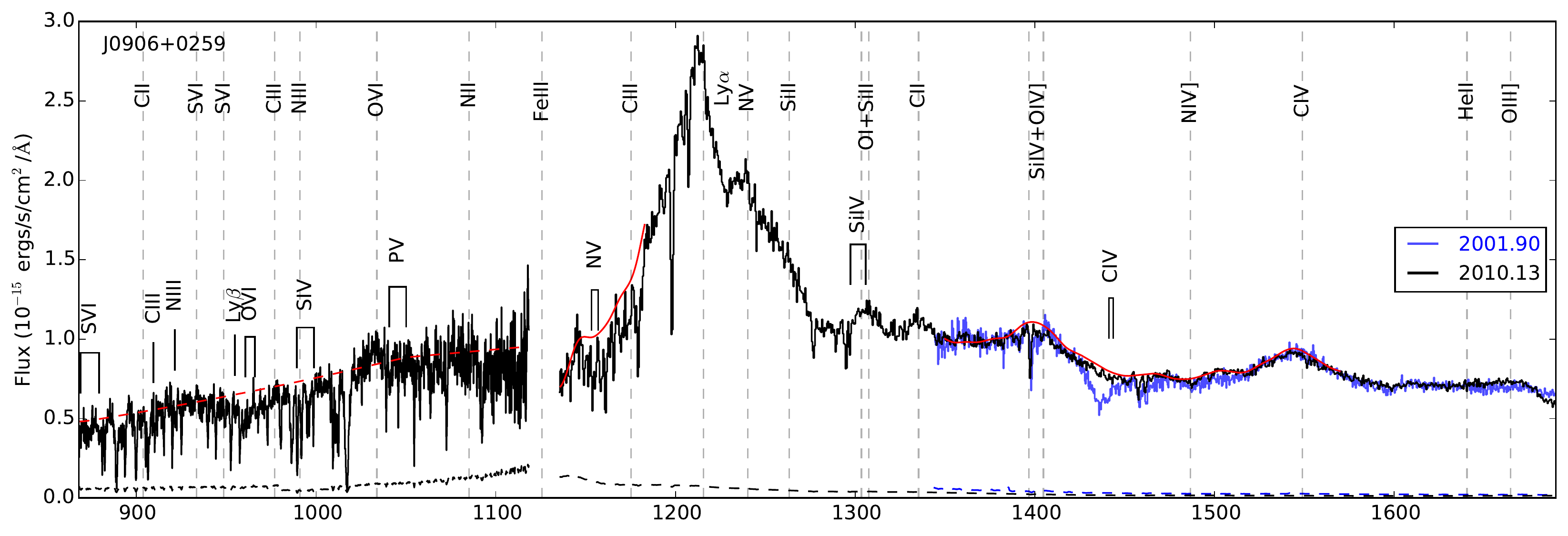}
 		\includegraphics[width=1.039\linewidth, height= 0.35\linewidth, keepaspectratio]{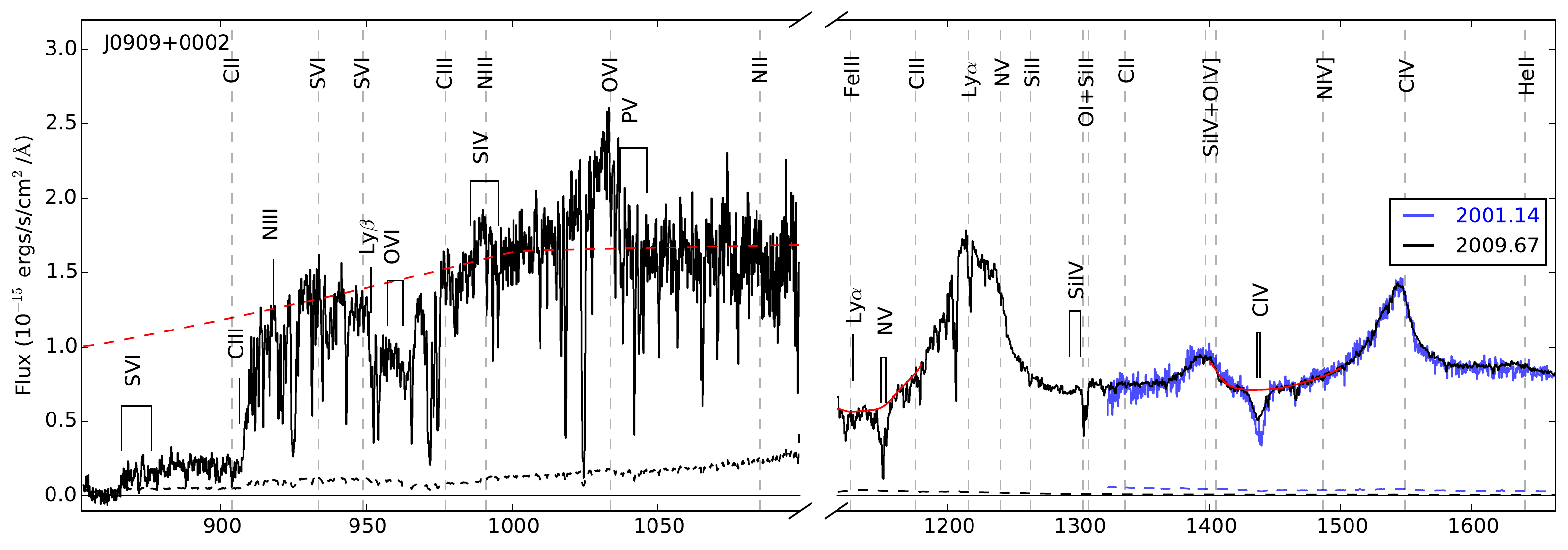}
 		\includegraphics[width=1.039\linewidth, height= 0.35\linewidth, keepaspectratio]{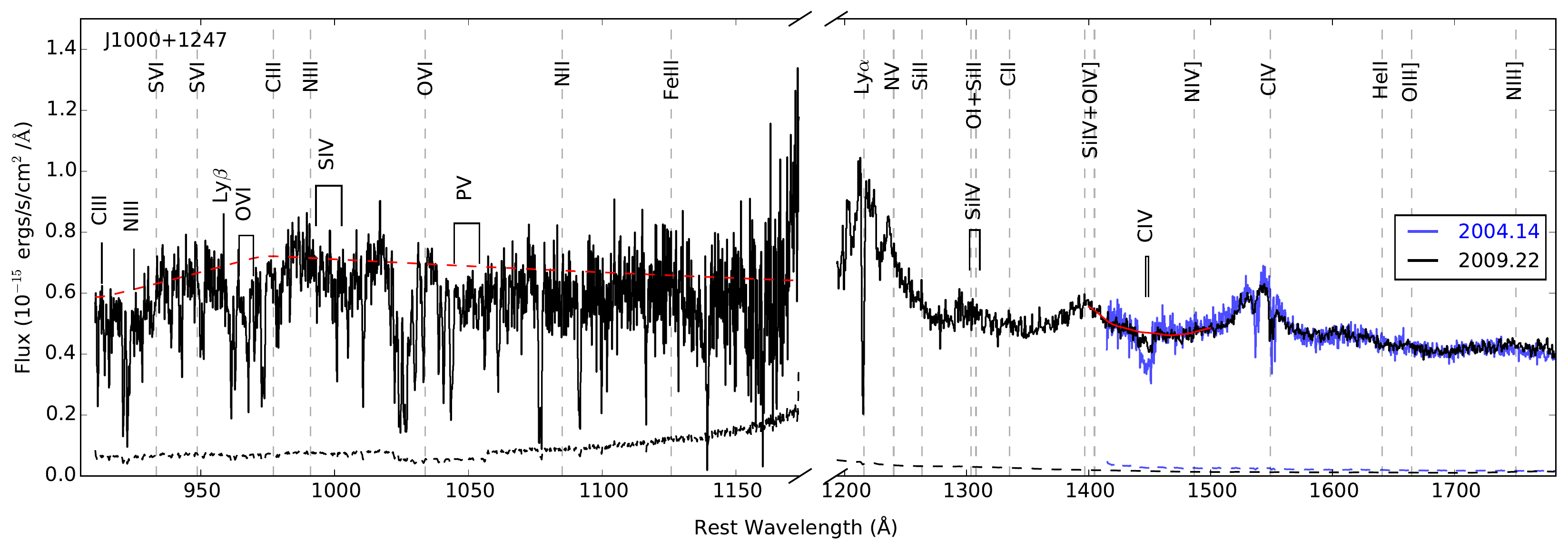}
 \caption{}
 \label{fig:hst_mdm}
	\end{minipage}
\end{figure*}
\begin{figure*}
\centering
	\begin{minipage}{\linewidth}
 		\includegraphics[width=\linewidth, height= 0.4\linewidth, keepaspectratio]{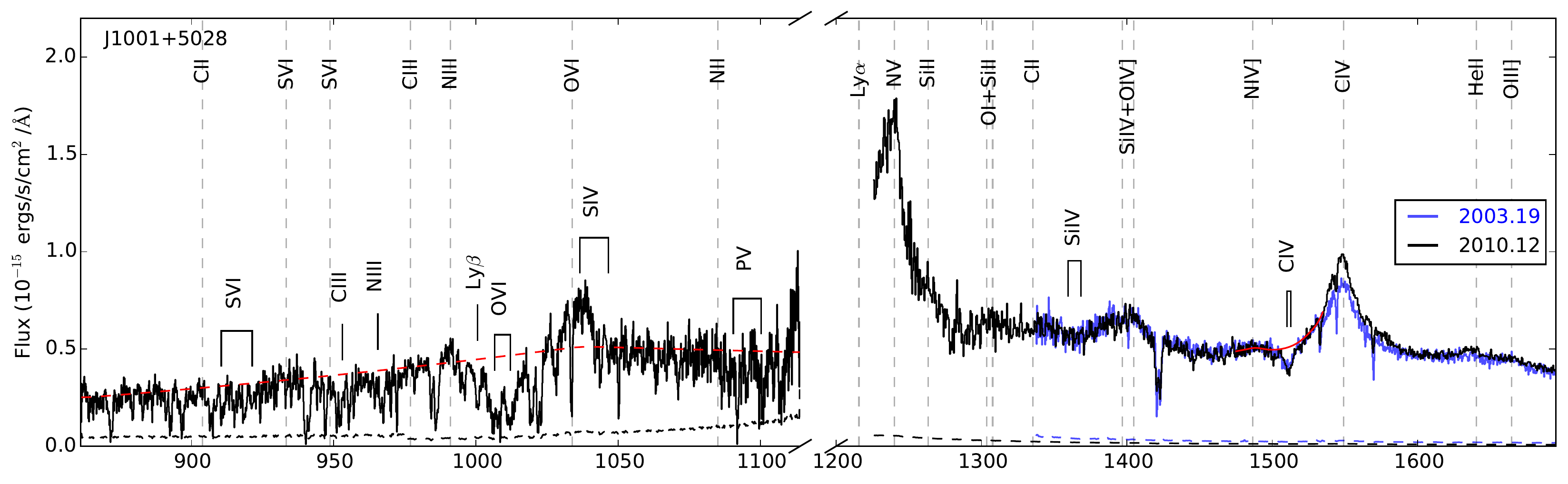} 
		\includegraphics[width=\linewidth, height= 0.4\linewidth, keepaspectratio]{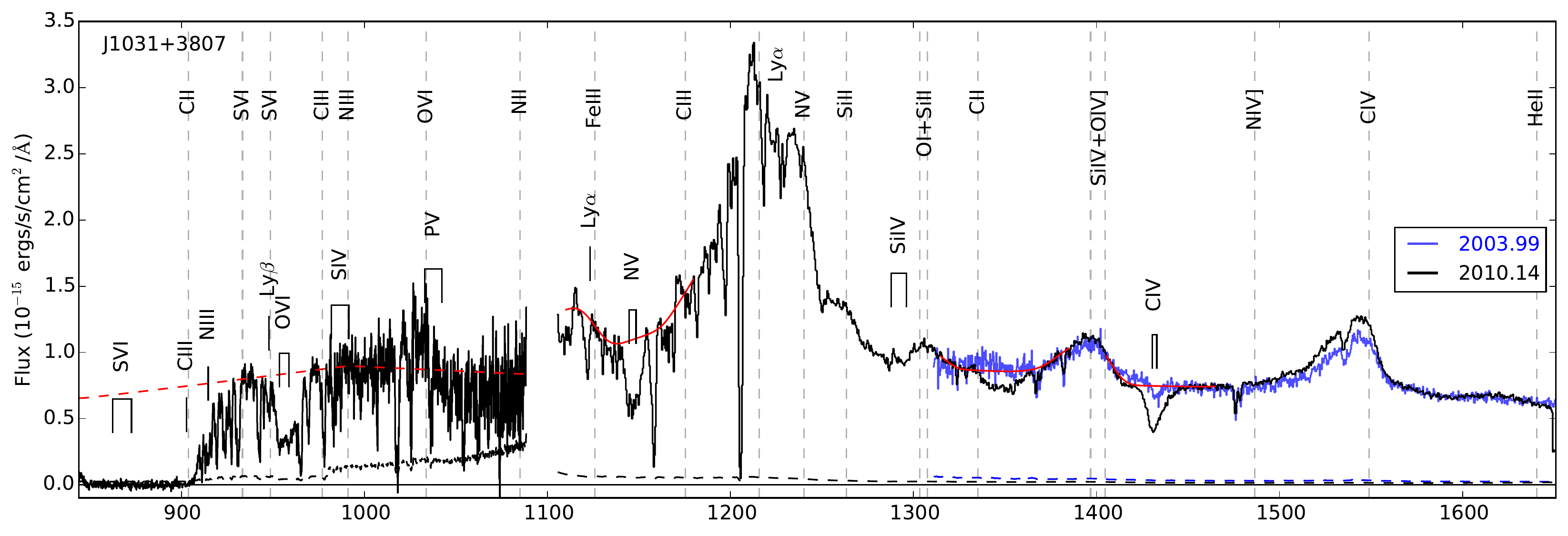}
		\includegraphics[width=\linewidth, height= 0.4\linewidth, keepaspectratio]{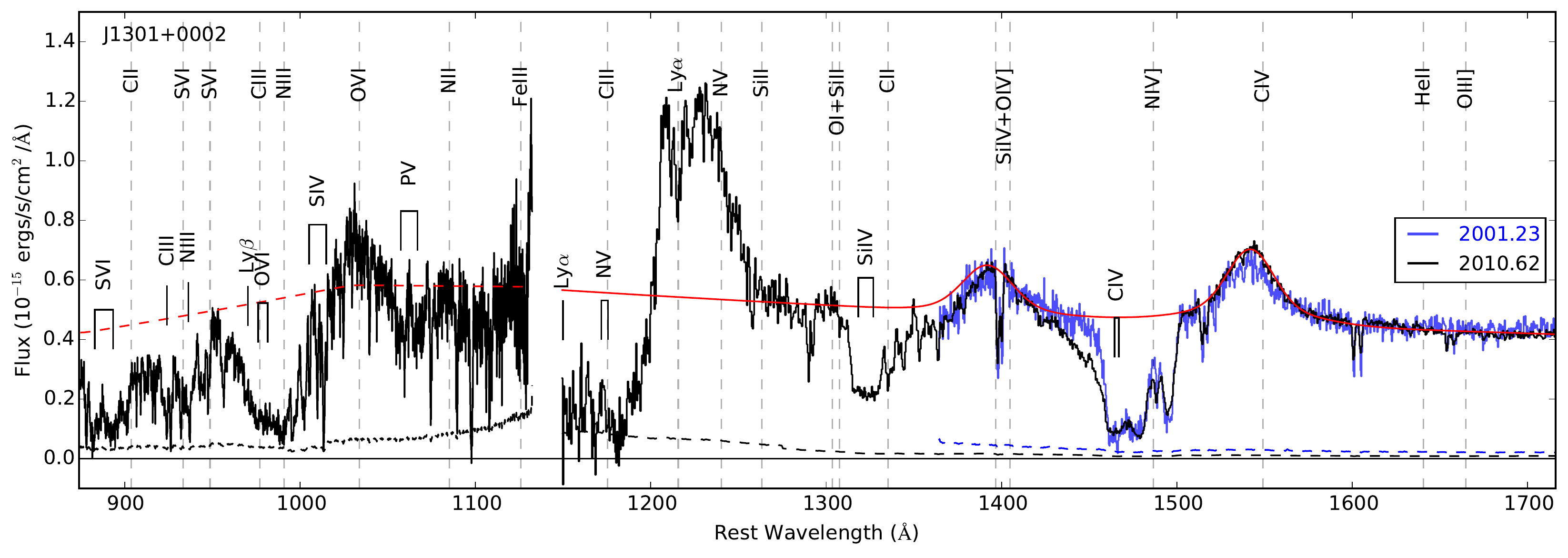}
\contcaption{SDSS and combined spectra of each quasar in the rest frame. The components of the combined spectrum are shown in black, while the SDSS spectrum used to select these quasars are plotted in blue (Obs1 in Table \ref{table:quasar_observations}). The HST and SDSS spectra are plotted without scaling, but the ground-based spectra are scaled to match SDSS (see \S\ref{measure} for details). The red dashed and solid lines show the fitted continua used to measure the absorption lines (see \S\ref{measure} for details). Absorption lines are marked above the spectra at positions expected from the ``initial" line fits (\S\ref{measure}). Common AGN emission lines are marked by dashed vertical lines. The error spectra are shown as dashed black and blue lines. We detect the emergence of a high-velocity BAL in J1031+3807 (see Figures \ref{fig:avg_flux} \& \ref{fig:103112_double_c4}). OVI is present in every quasar and stronger than CIV in all of the mini-BAL systems. J1301+0002 contains PV which indicates that CIV and other lines are severely saturated (see \S\ref{results}).}
\label{fig:continued}
	\end{minipage}
\end{figure*}

\begin{table*}
	\caption{Quasar Data Summary: A list of the observations of each quasar including specific things to note about certain observations such as epochs averaged together and which observations were contemporaneous with the HST observation. $\Delta t_{\mathrm{total}}$ is the time between the first and the last observation, $\Delta t_{\mathrm{shortest}}$ is the time between the most rapid variation, and $\Delta t_{\mathrm{largest}}$ is the time between the largest variation. All $\Delta t$'s are expressed in years in the rest frame of the quasar. The $\Delta A$'s are the change in the normalized amplitude of the variation (see \S\ref{variability} for more details).}
	\label{table:quasar_observations}
	\setlength{\tabcolsep}{3pt}
	\begin{tabular}{*{8}{c}}
	\hline
	Quasar & J0208+0022 & J0906+0259 & J0909+0002 & J1000+1247 & J1001+5028 & J1031+3807 & J1301+0002 \\
	\hline
	 Obs 1 & SDSS 2000.73 & SDSS 2001.90 & SDSS 2001.14 & SDSS 2004.14 & SDSS 2003.19 & SDSS 2003.99 & SDSS 2001.23 \\
	 Obs 2 & KPNO 2007.04 & MDM 2009.22 & KPNO 2006.34$^{\mathrm{A}}$ & MDM 2009.22$^{\mathrm{C}}$ & MDM 2009.22 & KPNO 2006.33$^{\mathrm{A}}$ & MDM 2009.22 \\
	 Obs 3 & MDM 2009.07 & MDM 2010.13$^{\mathrm{C}}$ & KPNO 2007.04$^{\mathrm{A}}$ & HST 2010.37 & KPNO 2010.12$^{\mathrm{C}}$ & KPNO 2007.4$^{\mathrm{A}}$& MDM 2010.13$^{\mathrm{A}}$\\
	 Obs 4 & HST 2010.03 & HST 2010.21 & KPNO 2007.22$^{\mathrm{A}}$ & MDM 2013.18 & HST 2010.45 & MDM 2009.22 & HST 2010.61\\
	 Obs 5 & MDM 2010.04$^{\mathrm{C}}$ & MDM 2013.19 & MDM 2009.22$^{\mathrm{B}}$ & - & MDM 2013.18 &  MDM 2010.14$^{\mathrm{C}}$ & BOSS 2011.11$^{\mathrm{A}}$ \\
	 Obs 6 & - & - & KPNO 2010.11$^{\mathrm{B}}$  & - & BOSS 2014.01 & HST 2010.49 & MDM 2013.19 \\
	 Obs 7 & - & - & HST 2010.21 & - & - & MDM 2013.19 & - \\
	 Obs 8 & - & - & BOSS 2010.92 & - & - & - & - \\
	 Obs 9 & - & - & MDM 2013.19 & - & - & - & - \\
	 AvgA & - & & Avg 2006.87 & - & - & Avg 2006.69 & Avg 2010.62$^{\mathrm{C}}$ \\
	 AvgB & - & - & Avg 2009.67$^{\mathrm{C}}$ & - & - & - & - \\
	 $\Delta t_{\mathrm{total}}$ & 3.21 & 3.99 & 4.19 & 3.36 & 3.81 & 3.18 & 4.29 \\
	 $\Delta t_{\mathrm{shortest}}$ & 0.70 (2-3) & 1.08 (3-4) & 0.69 (2-3)& 1.89 (1-2) & 0.29 (4-5) & 1.05 (2-3) & 0.33 (2-3) \\
	 $\Delta A_{\mathrm{shortest}}$ & -0.087$\pm$0.015 & -0.057 $\pm$ 0.009 & -0.046 $\pm$ 0.007 & -0.14 $\pm$ 0.014 & 0.063 $\pm$ 0.008 & -0.057 $\pm$ 0.008 & 0.039 $\pm$ 0.008 \\ 
	 $\Delta t_{\mathrm{largest}}$ & 0.70 (2-3) & 3.99 (1-4) & 3.4 (1-4) & 3.36 (1-3) &3.81 (1-5) & 2.12 (1-3) & 4.29 (1-4) \\ 
	 $\Delta A_{\mathrm{largest}}$ & 0.087$\pm$0.015 &-0.25 $\pm$ 0.011 &-0.18 $\pm$ 0.011 & -0.16 $\pm$ 0.016 & 0.21 $\pm$ 0.013 & 0.33 $\pm$ 0.011 & 0.38 $\pm$0.018 \\
	\hline
	\multicolumn{8}{l}{$^{\mathrm{A}}$ Epochs averaged together to make AvgA.}\\
	\multicolumn{8}{l}{$^{\mathrm{B}}$ Epochs averaged together to make AvgB.} \\
	\multicolumn{8}{l}{$^{\mathrm{C}}$ The ground-based epoch that was roughly contemporaneous with the HST observation.} \\
	\end{tabular}
\end{table*}

The quasars in our study are from the catalog of CIV $\lambda$1548,1551 $\AA$ absorption lines created by \cite{Nestor08} and \cite{Paola09}. This catalog contains the brightest $\sim$2200 quasars at redshifts $z_{em}$ $\gtrsim$ 1.6 in the Sloan Digital Sky Survey (SDSS) Data Release 4 (DR4) (\citealt{Abazajian09}, \citealt{Schneider10}). CIV absorption lines were identified and measured in the spectra using an automated fitting procedure described by \cite{Nestor08}. Broad outflow lines in this sample (e.g., BALs and mini-BALs) were flagged and confirmed by visual inspection. The mini-BALs and similar broad outflow lines were reanalysed to carefully determine their rest equivalent widths (REWs), full widths at half minimum (FWHMs), and velocity shifts from quasar emission line redshifts (\citealt{Paola09}). 

We select seven bright quasars that span a range in strengths and widths in the CIV outflow lines from weak/narrow mini-BALs to one strong/broad BAL. Table \ref{table:quasar_properties} lists some basic properties of the quasars and their outflow lines derived from their SDSS spectra. This includes the emission-line redshifts \citep[from][]{Hewett10} and the outflow line balnicity indices (BI), FWHMs, and the absorption line redshift (from \citealt{Paola09}). Only the FWHM for J1301+0002\footnote{We use abridged quasar names like this throughout the paper. The full SDSS quasar names are listed in Table \ref{table:quasar_properties} .} was measured by us from the SDSS spectrum. Several of the outflow lines are marginal BALs with 0 $<$ BI $<$ 300 km/s (see \citealt{Weymann91}, also \citealt{Dawson13}, \citealt{Paris14}). For convenience, we will refer to all of the features with BI $<$ 300 km/s hereafter as mini-BALs. 

The quasars are also required to have redshifts in the narrow range $z_{em}\sim 1.6$--1.8 in order to i) place important diagnostic lines such as  SVI 933,944 $\AA$, OVI 1032,1038 $\AA$ and PV 1118,1128 $\AA$ in the sensitive range of the COS NUV detectors, and ii) minimize contamination by the Ly$\alpha$ forest at wavelengths below 1216 $\AA$. The last column in Table \ref{table:quasar_properties} lists the quasar bolometric luminosities. These are estimated from the observed fluxes at 1450 $\AA$ in the SDSS spectra (see \S\ref{data} below) and a bolometric correction given by $L = 4.4\,\lambda L_{\lambda}$(1450 $\AA$) (\citealt{Hamann11}) in a cosmology with $H_0$ = 69.6 km s$^{-1}$, $\Omega_M$ = 0.286, and $\Omega_{\Lambda}$ = 0.714 \citep{Wright06}. 

J1001+5028 is a known micro-lensed quasar (\citealt{Oguri05}, \citealt{Hales07}). According to \cite{Oguri05}, there is a time delay of $\Delta t \sim 45 h^{-1}$ days between the two source images. In \cite{Oguri05}, their A image corresponds to the SDSS and MDM spectra in this study, while their B image corresponds to the BOSS spectrum in this study. The absorption lines at $\sim$1400 \AA\ in the BOSS spectrum have a separation of  $\sim$754 $\pm \sim$50 km s$^{-1}$ which encompasses the published value of MgII of 770 km s$^{-1}$. Since the published value is within our error, we conclude these lines are indeed MgII which could indicate that this is part of the lens.

\section{Data Sets \& New Observations} \label{data}
Table \ref{table:instrum} lists basic information about the telescopes and instruments used to obtain spectra for this study. Table \ref{table:quasar_observations} gives more details about the observations such as dates and number of observations. 

We supplement the SDSS spectra described in \S\ref{sample} with new observations of each quasar. First, we observed each quasar once in the rest-frame far-UV using HST COS/NUV. Second, we obtained ground-based spectra at various other epochs to study the variability in the CIV outflow lines in the rest-frame near-UV using the Kitt Peak National Observatory (KPNO) 2.1m and the MDM Observatory 2.4m telescopes. The new observing programs were designed such that one ground-based observation was roughly contemporaneous (within a few months) with that of the HST observation (see Table \ref{table:quasar_observations} for details). Hereafter, we refer to the combination of the HST spectrum and its near-simultaneous ground-based spectrum as the ``combined spectrum" for each quasar. These combined spectra are shown in Figure \ref{fig:hst_mdm}, along with the SDSS spectra used to initially measure the outflow lines and select quasars for this study (\S\ref{sample}). In addition to these new observations, we searched the Baryon Oscillation Spectroscopic Survey of SDSS-III (BOSS, \citealt{Dawson13}, \citealt{Paris14}, \citealt{Alam15}) archives for additional epochs of data. 

For HST COS, we used the G230L grating at two central wavelengths, 2636 $\AA$ and 2950 $\AA$ with a wavelength coverage per setting of roughly $\sim$400 $\AA$. The spectra obtained in the two grating setting could have flux mismatches up to 17\% in the $\sim$100 $\AA$ region of overlap. To rectify this mismatch, we scaled the long-wavelength spectra to match the short-wavelength spectra using a ratio of the median fluxes in the overlap region. We then combined the spectra in the overlap region using an inverse variance weighted average. In these HST spectra, the Lyman limits that appear in the rest frame of J0909+0002 and J1031+3807 are not related to the outflow lines that we study here in this paper.

The spectra provided from the HST pipeline were noisier than expected. We attribute this to the large extraction aperture used by the pipeline which resulted in large contributions from background/sky emissions. We therefore performed our own extractions\footnote{We performed these extractions using a code Martin Durant developed in the Python programming language.} with a smaller extraction box. We adopted a box size of 15 pixels based on trial and error experimentation. This box size was a compromise between smaller boxes desirable for reduced backgrounds and noise, and larger boxes that include more flux in the quasar point spread function.

Most of our new ground-based spectra were acquired between 2009 and 2013 using the MDM Observatory 2.4-m Hiltner telescope with the Ohio State University Boller and Chivens CCD spectrograph (CCDS\footnote{http://www.astronomy.ohio-state.edu/MDM/CCDS/}). Since CCDS has a narrow wavelength range of $\sim$1600 $\AA$, we used different settings for each quasar which were dictated by the redshift of the quasar in order to cover at least the rest wavelength range of Ly$\alpha$ emission to CIV emission (\citealt{Capellupo11}, \citeyear{Capellupo12}). Additional spectra were obtained in 2006 - 2007 using the 2.1m telescope at KPNO using the Goldcam spectrometer. The ground-based data were reduced using standard procedures (bias subtraction, flat-fielding, wavelength and {\it relative} flux calibrations) using the IRAF software package. For more details on the MDM data and its reduction see \cite{McGraw15} and \cite{Capellupo11}. Specifics for the KPNO data and its reduction are presented in \cite{Paola09}. Absolute flux calibrations were not performed on any of these data due to weather and time constraints. 

The first nine rows in Table \ref{table:quasar_observations}, labeled Obs1 - Obs9, indicate the telescopes and dates of all of the individual observations. The next two rows, AvgA and AvgB, identify several average spectra we created to improve the signal-to-noise ratio. The individual spectra used in these averages are noted with A and B superscripts. Average spectra were created for the few cases where the individual observations had no evident changes between the spectra and were close together in time. These averages use inverse-variance weights with the pixel-by-pixel variance determined from the error spectra after rebinning to a common wavelength grid. For the KPNO data where error spectra are not available, we estimate the variance from pixel-to-pixel fluctuations in the flux spectra. Notice that the combined spectra of quasars J0909+0002 and J1301+0002 use one of these averages (Figures \ref{fig:hst_mdm}). Other parameters listed near the bottom of Table \ref{table:quasar_observations} pertain to our CIV variability analysis described in \S\ref{variability} below.  

\section{Analysis \& Results} \label{results}
We fit all of the broad outflow lines detected in the near-simultaneous combined spectra of each quasar shown in Figure \ref{fig:hst_mdm}. This includes rest wavelengths from at least $\sim$915 $\AA$ to $\sim$1750 $\AA$. We do not measure unresolved narrow CIV systems that are also present in some of the quasars. \S\ref{measure} describes the line fits while \S\ref{variability} assesses the CIV line variabilities based on our multi-epoch ground-based monitoring.

\subsection{Line Fits and Identification}\label{measure}
Before fitting the lines, we first define the unabsorbed continuum that includes the quasar continuum spectrum plus any broad emission lines at the absorption line wavelengths. We fit the HST spectra with two power laws (of the form $F_{\lambda}\sim\lambda^{\alpha}$) on either side of crudely near 1000 $\AA$ to accommodate a spectral break that is evident near this wavelength, and we constrain these power laws by the observed flux in narrow wavelength windows that appear to be free of absorption and emission lines based on visual inspection. We specifically avoid absorption lines in the Ly$\alpha$ forest, and we avoid wavelengths that {\it might} have weak emission lines based on previous observations of low-redshift AGN (\citealt{Shull12}, \citealt{Stevans14}). The results are shown by the dashed red curves on the left side of Figure \ref{fig:hst_mdm}.

For the ground-based spectra, we use tools in the IRAF task SPLOT to manually interpolate a continuum across the CIV absorption lines. We then smooth this interpolated spectrum with a Gaussian filter to create a pseudo-continuum that accounts for all sources of emission in the unabsorbed spectrum. The one exception was that of J0906+0259 where the unabsorbed continuum was difficult to determine due to CIV low level absorption across a wide velocity range. In this case, we smoothed the epoch with the least CIV absorption (MDM 2013.19) to create the continuum. These continua are shown by the solid red curves in Figure \ref{fig:hst_mdm}. 

The BAL quasar, J1301+0002, is a special case because of complications caused by the BALs (see Figure \ref{fig:hst_mdm}). This is particularly problematic at wavelengths where multiple BAL troughs blend together to suppress the flux at all (or nearly all) wavelengths. In the HST BAL spectrum, we again use two power laws but with a break beneath the OVI emission line instead of 1000 $\AA$. The longer wavelength segment is forced to match unabsorbed spectral regions roughly as before but with larger uncertainties due to weak BAL contributions. The short-wavelength power law is more problematic because multiple BAL troughs blend together at wavelengths below OVI 1032,1038 $\AA$. It is not clear that any part of this spectrum represents the true quasar continuum. We anchor the short-wavelength power law to the highest point in the observed spectrum near 950 $\AA$. This yields roughly the lowest possible continuum consistent with the data and, therefore, conservative lower limits on the BAL strengths measured at these wavelengths. In the ground-based BAL spectrum, we fit a single power law constrained by the median flux in three spectral windows that avoid strong emission and absorption lines, namely 1365-1390, 1520-1540, and 1610-1675 $\AA$. Then we add Gaussians to represent the CIV and SiIV emission lines unaffected by BAL absorption. Two gaussians are used for each line and the SiIV profile is forced to have the same redshift as CIV. The continuum results for both the HST and ground-based spectra are again shown by the dashed and solid red curves, respectively, in Figure \ref{fig:hst_mdm}. A slight mismatch between the data and the continuum on the blue side of the SiIV emission line is caused by the forced redshift match to CIV. We do not make adjustments because the continuum in this region has no bearing on our analysis.

Next, we fit the BALs and mini-BALs assuming that the absorbing regions are spatially homogeneous in front of a spatially uniform emission source. The measured line intensities at each wavelength are then given by 
\begin{equation}\label{line_fit}
	I_{\lambda} \ = \ I_0 (1-C_{\lambda}) + C_{\lambda} I_0 e^{-\tau_\lambda}
\end{equation}
where $I_0$ is the unabsorbed continuum intensity, $C_{\lambda}$ is the line-of-sight coverage fraction of the absorbing material in front of the emission source (with $0\leq C_{\lambda} \leq 1$), and $\tau_\lambda$ is the line optical depth (\citealt{Hamann97}, \citealt{Hamann04}). A standard technique for deriving values of both $C_{\lambda}$ and $\tau_{\lambda}$ is to fit the lines in doublets simultaneously, such as CIV 1548,1551 $\AA$ and OVI 1032,1038 $\AA$, using their known optical depth ratios of $\sim$2:1. This provides two equations to solve for the two unknowns. The main caveat is that the geometry can be more complex. We will see below that, for the BAL quasar J1301+0002, different ions/lines have different covering fractions and therefore the absorber is not spatially homogeneous. Nonetheless, Equation \ref{line_fit} provides useful constraints on $C_{\lambda}$ and $\tau_{\lambda}$, especially in limiting cases as in the present study where the lines appear to be optically thick or we do not have resolved doublet ratios to determine $C_{\lambda}$ and $\tau_{\lambda}$ separately (see \citealt{Hamann04}, \citealt{Hamann11} and refs. therein for more discussion of the results and assumptions of this analysis). 

For simplicity, our line fits also assume that the covering fraction is constant across the line profiles. Thus, we refer only to this constant covering fraction as $C_0$ hereafter. Finally, we adopt Gaussian optical depth profiles, which leads to a simple expression for the absorber column densities given by 
\begin{equation}
 	\label{nion}
	N_{\mathrm{ion}} \ = \ 6.68 x 10^{14} \left(\frac{b\,\tau_0}{f\lambda_0}\right) ~ ~ ~  \mathrm{cm^{-2}}
\end{equation}
where $f$ is the line oscillator strength, $\lambda_{0}$ is the laboratory wavelength (in \AA ), $b$ is the Doppler parameter (in km s$^{-1}$), and $\tau_0$ is the line-centre optical depth. For doublets such as CIV and OVI, we use two gaussian components (even if they are blended in the observed spectrum) forced to have the same $b$ value and velocity shift. The velocity shifts are measured relative to the quasar emission-line redshifts listed in Table \ref{table:quasar_properties}. 

Our line procedure for the different lines in each quasar follows several steps, beginning with an ``initial fit'' that we use as a template to fit other lines in the same system. The procedure is slightly different for the BAL and mini-BAL quasars. One notable difference is that the procedure we use to define a continuum for the mini-BAL quasars works very well (much better across the mini-BAL features than any simple fitting procedure that might combine power laws with gaussian emission lines). However, this procedure for determining a continuum for the mini-BALs does not work for the BALs because the BAL features are too broad. One needs to interpolate extensively across the BAL troughs guided by some fit assumptions. Since we want to understand the structure of the outflow, we only look at systems that have unifying properties such as redshift and Doppler parameter ($b$).

\subsubsection{Mini-BAL Quasars}\label{mini_fits}

\begin{figure*}
\begin{center}
 \includegraphics[scale=0.5,angle=-0.0]{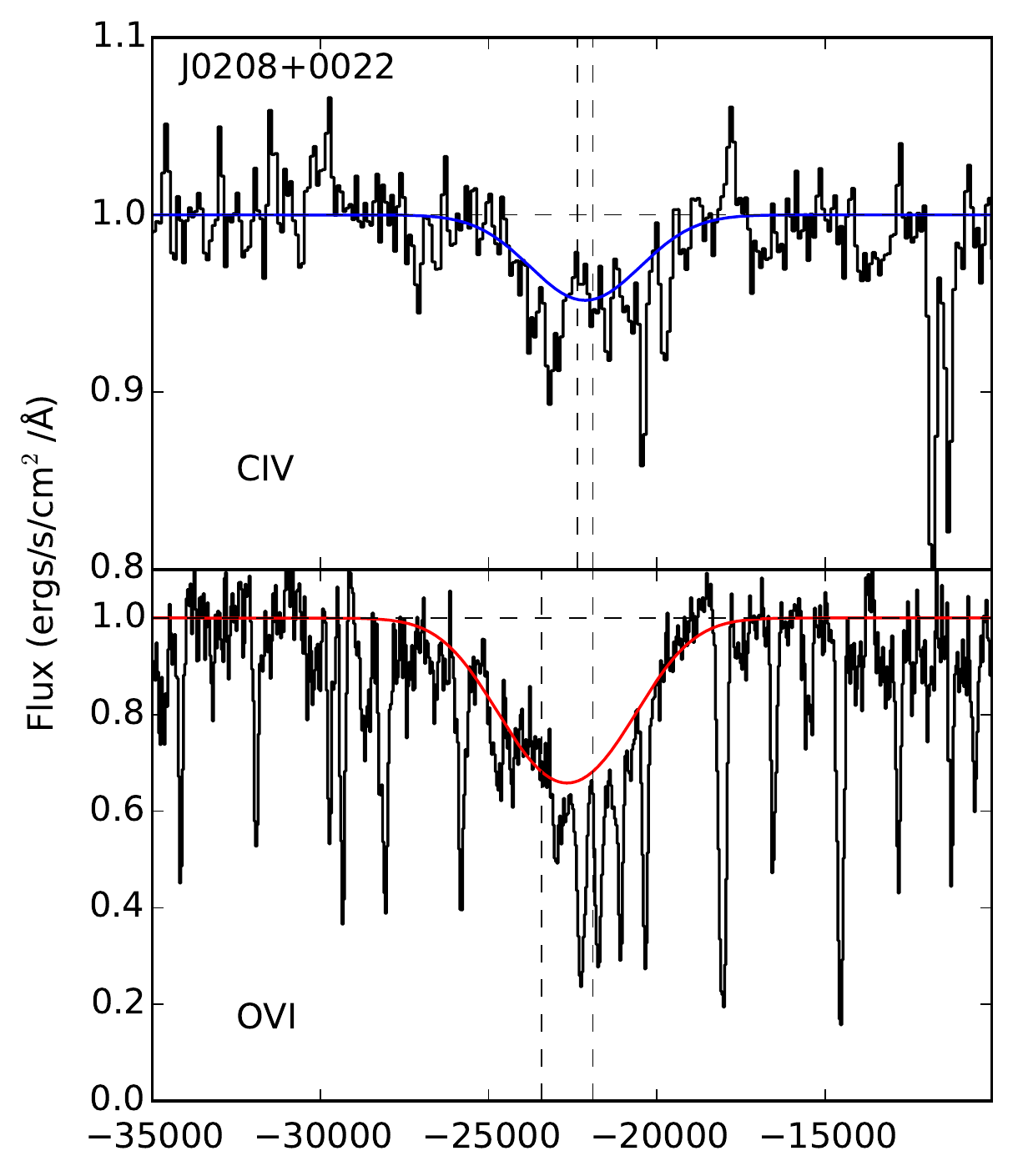}
 \includegraphics[scale=0.5,angle=-0.0]{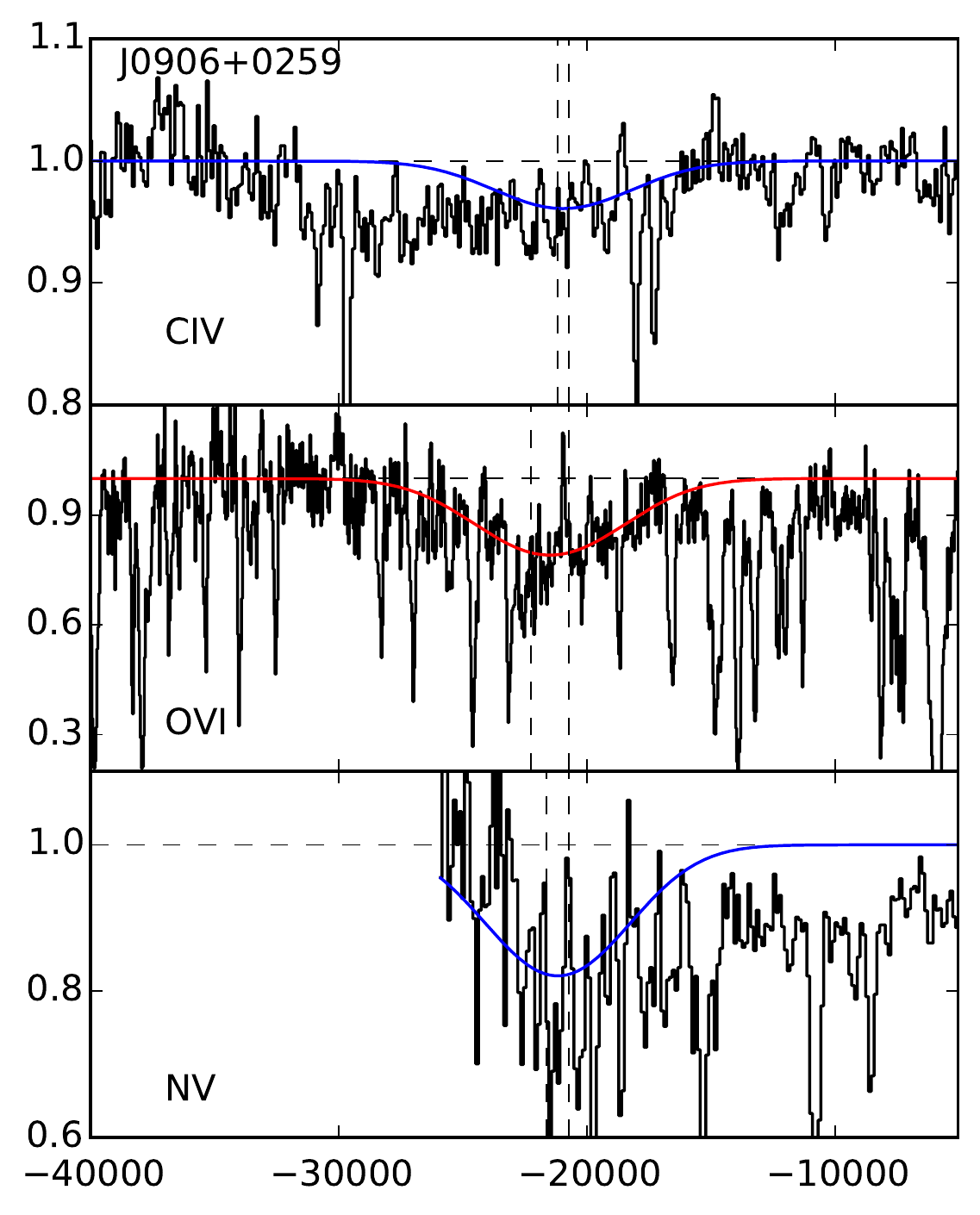}
 \includegraphics[scale=0.5,angle=-0.0]{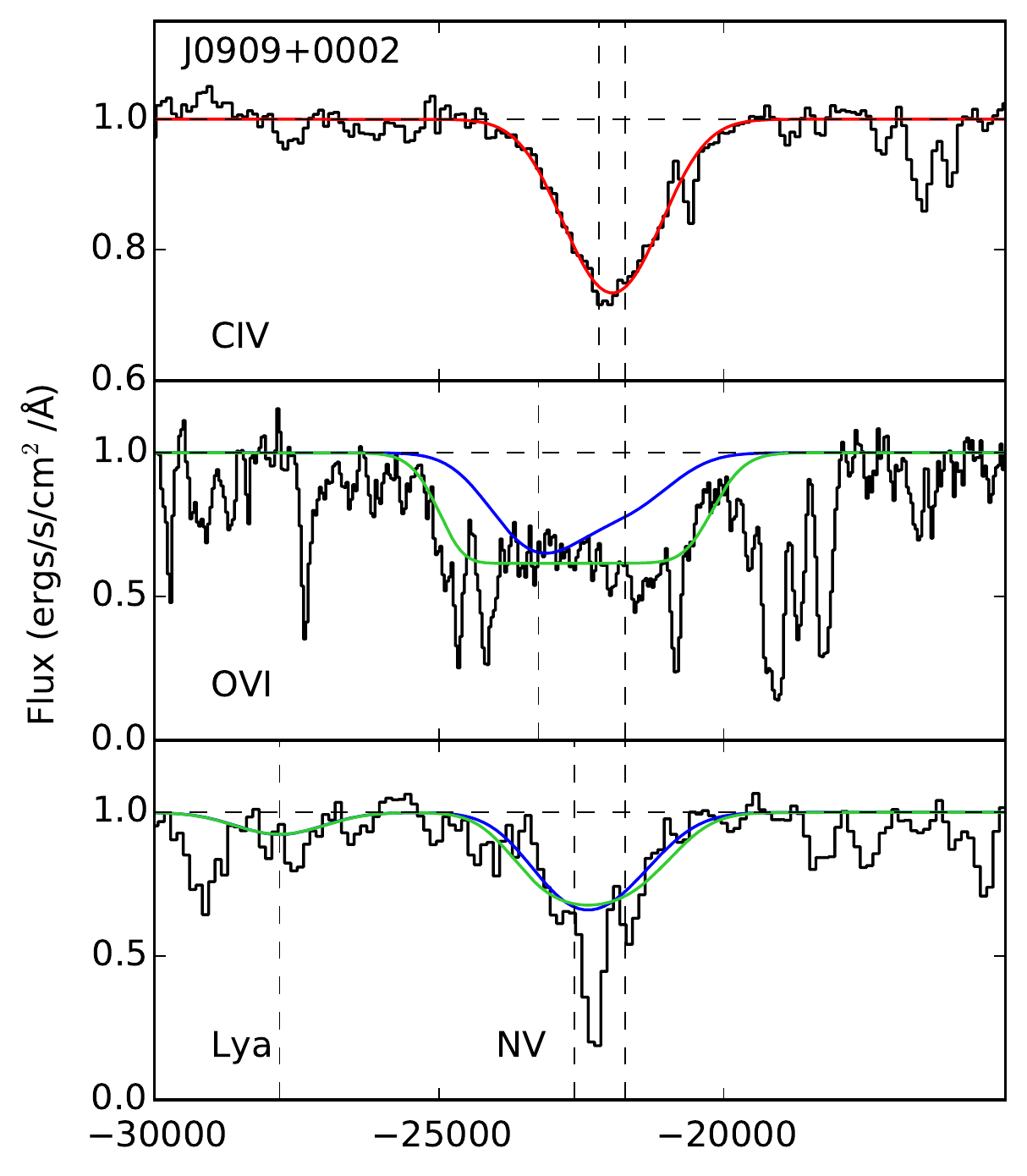}
 \includegraphics[scale=0.5,angle=-0.0]{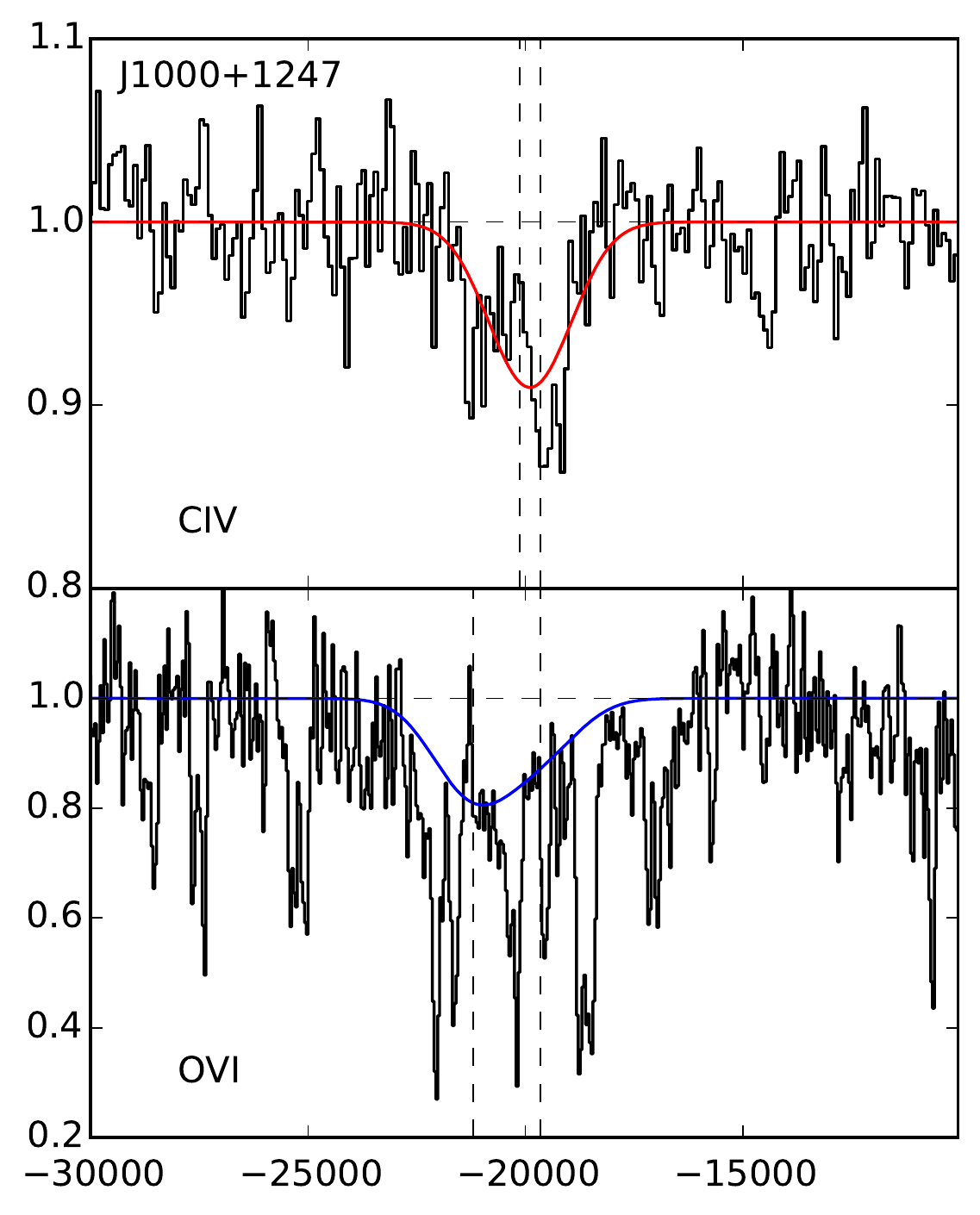}
 \includegraphics[scale=0.5,angle=-0.0]{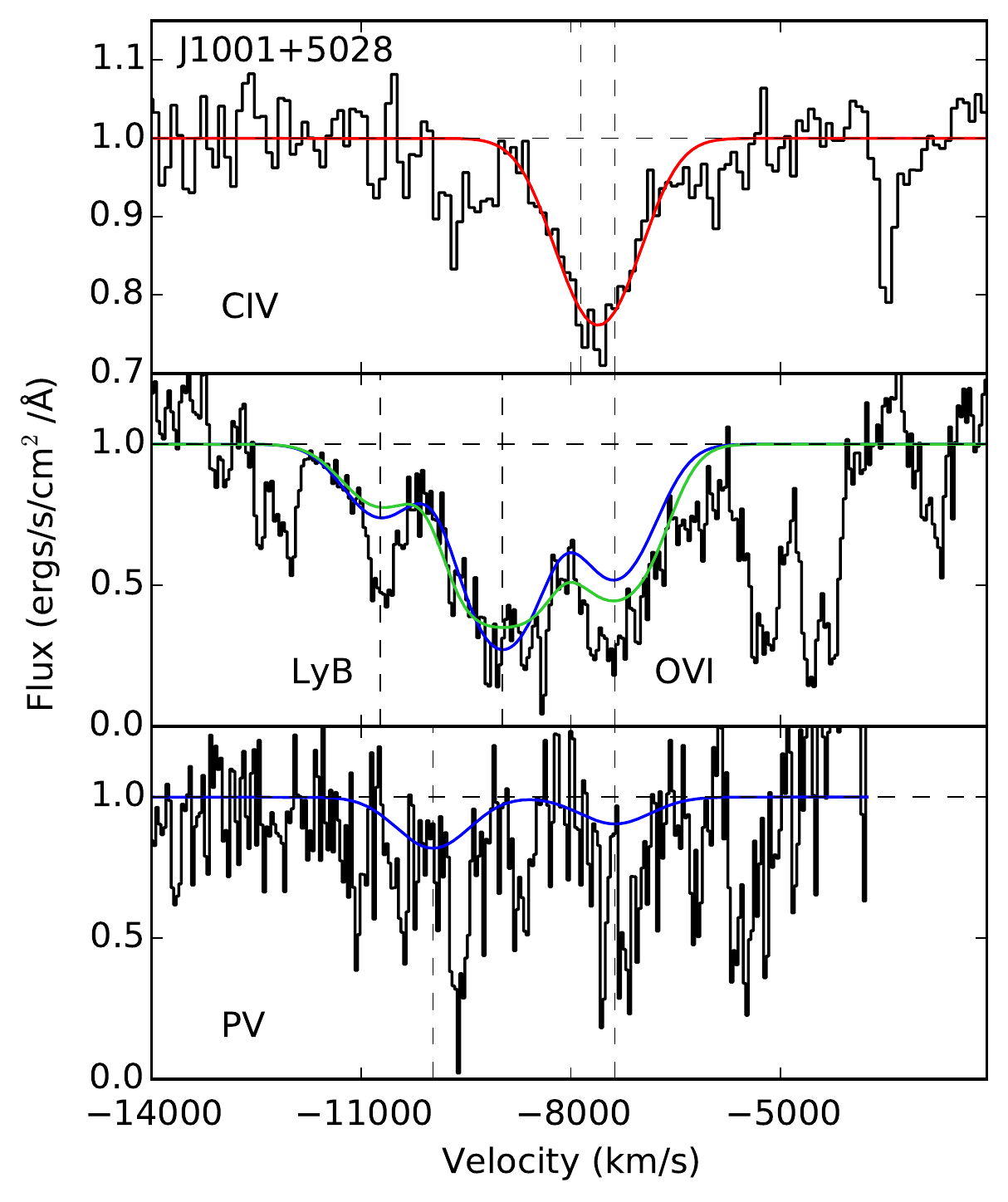}
 \includegraphics[scale=0.5,angle=-0.0]{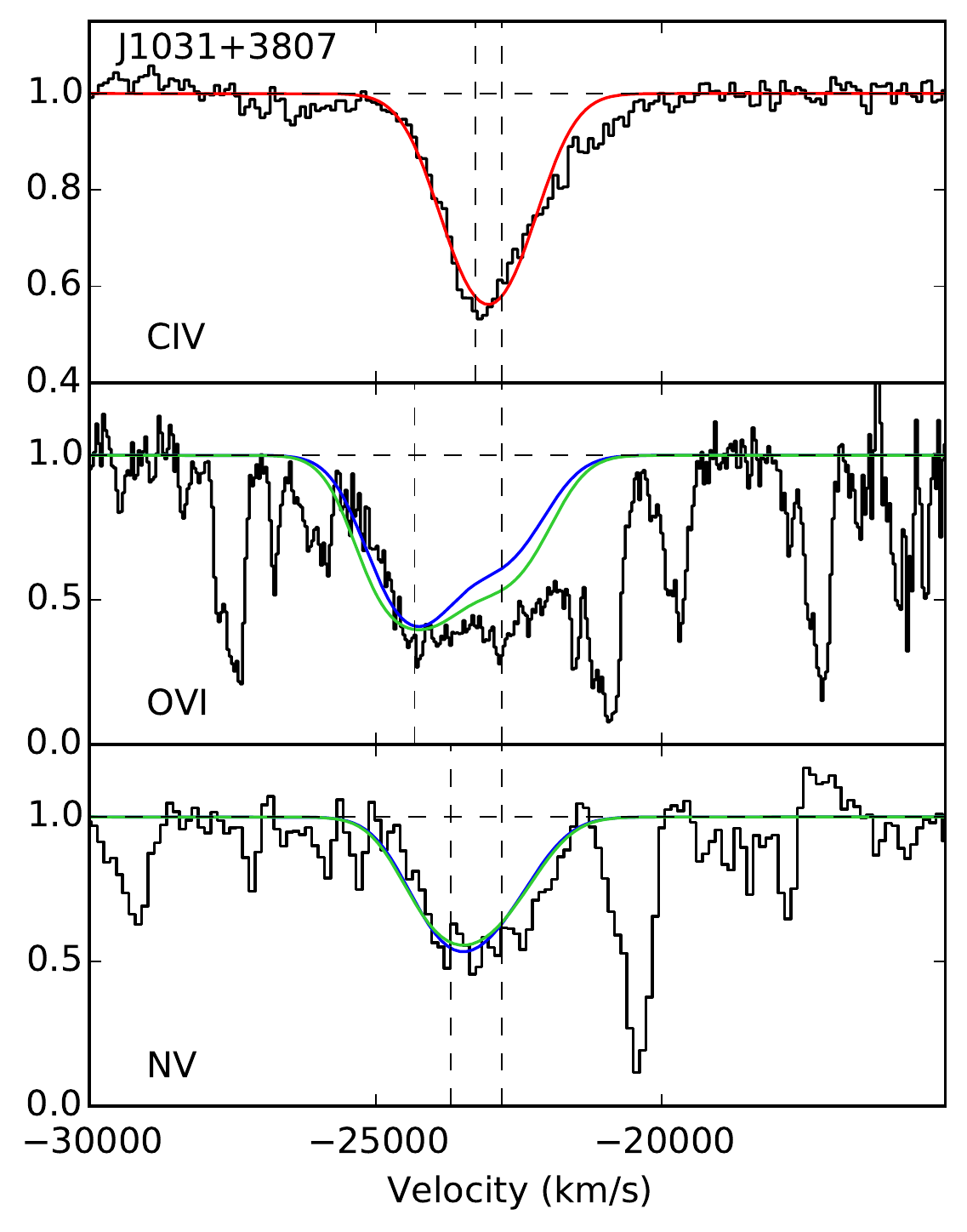}
 \end{center}
 \vspace{-12pt}
 \caption{Fits to the absorption lines in the normalized combined spectra of the six mini-BAL quasars. The flux is normalized as described in \S \ref{measure}. The velocities are relative to the quasar emission-line redshift (Table \ref{table:quasar_properties}) for the longer wavelength components in doublet lines. For each quasar, the red curves are the initial fits (\S\ref{mini_fits}) which define basic parameters such as velocity shift and doppler parameter for the gas systems and have $C_0$ = 1 as they are unresolved doublets. The blue curves are the initial fits moved to a new ion with different and visually determined $\tau_0$ values. The lime green curves are again the initial fits, but have visually determined $C_0$ < 1 and $\tau_0$. NV, PV, and Ly$\alpha$ green fits have the $C_0$ of OVI. J0909+0002, J1001+5028, and J1031+3807 all have indications of partial covering from saturated ($\sim$1:1) OVI doublets that do not reach zero intensity.}
 \label{fig:fits}
\end{figure*}

We fit the mini-BALs in the combined spectra as follows:

\begin{enumerate}
	\item \textit{Initial Fit:} First, we fit one strong and (relatively) uncontaminated line/doublet to obtain basic line profile parameters $b$ and the velocity shift. The line selected for this initial fit is usually CIV as it is not contaminated by blends in the Ly$\alpha$ forest. However, in two quasars (J0208+0022 and J0906+0259) the CIV mini-BAL originally present in the SDSS spectrum weakened considerably to become less obvious in the combined spectrum measured several years later. In those cases, we obtain a more reliable initial fit from OVI. Since the doublets are unresolved in all of these initial fits, we do not have direct constraints on the line optical depths and covering fractions from the doublet ratios. Thus, we adopt $C_0 = 1$ and we use an artificial doublet optical depth ratio of 1:1 because this yields symmetric doublet profiles for any $\tau_0$ that are a good match to the generally-symmetric mini-BALs in the data. (We will argue below that many of the lines actually have $C_0 < 1$ and $\tau_0 > 1$ in spite of their shallow appearance in the spectra.) These initial fits yield firm lower limits on the line optical depths and ionic column densities, and they provide profile parameters useful for fitting other lines in the spectra, as described below. They are shown by the red curves in Figure \ref{fig:fits} with derived line properties listed in Table \ref{table:lines} (described further below). 
	\item Second, we visually inspect each spectrum for other outflow lines at the same velocity as the initial fit. We search specifically for lines detected previously in BAL and mini-BAL quasars, including OVI 1032,1038 $\AA$, PV 1118,1128 $\AA$, NV 1239,1243 $\AA$ etc. (e.g., \citealt{Leighly09}, \citealt{Paola11}, \citealt{Baskin13}, \citealt{Capellupo14}, Herbst in prep.). Then we fit each detected line/doublet using the initial fit as a template and manually scale only $\tau_0$ to match the data while avoiding unrelated absorption features (e.g., in the Ly$\alpha$ forest). We use physical optical depth ratios of 2:1 for the common doublets but again adopt $C_0=1$ to again derive lower limits on $\tau_0$ and the column densities. These fits are shown by blue curves in Figure \ref{fig:fits}. 
	\item Third, in some of the spectra (e.g., J0909+0002, J1001+5028, and J1031+3807) the OVI doublets are sufficiently resolved to indicate $\sim$1:1 doublet ratios and therefore optically thick absorption with $C_0<1$. In these cases, we perform additional fits by using the Doppler parameter (b) from the initial fit to CIV and manually adjust $\tau_0$ and $C_0$ to get the best fit to the data across both doublet components producing more realistic estimates of $C_0$, $\tau_0$ and the column densities. These fits are shown by the green curves in Figure \ref{fig:fits}. The procedure we adopt is to estimate $C_0$ from the measured depths of these saturated troughs, and then scale $\tau_0$ to match the data (ignoring obvious unrelated Ly$\alpha$ forest lines) while not allowing the fit to go significantly below the data in any part of the line profiles. This last constraint, combined with the fixed b value adopted from the previous fitting step, results in fitted lines in J1031+3807 that have only moderate optical depths (not yielding a $sim$1:1 doublet ratio). All of these fits are only illustrative, to show the limits obtained for the adopted b values. The general result from the $\sim$1:1 appearance of these OVI doublets in the data is that they have $\tau_0$ $\ge$ 3. 
	\item Fourth, for the three quasars mentioned above in step iii, where OVI provides evidence for $C_0<1$, we use the $C_0$ results to fit other lines such as NV and CIV in the same system. These fits with $C_0<1$ are not shown for CIV (because they look the same as the previous fits), but for other lines these fits are again shown by green curves in Figure \ref{fig:fits}. These $C_0<1$ fits yield more realistic lower limits on $\tau_0$ and the ionic column densities and are the ``final fits" in these cases.
\end{enumerate}

Table \ref{table:lines} lists the line parameters and column densities inferred from the fits shown in Figures \ref{fig:fits} and \ref{fig:bal_fits}. For the doublets, the REW is the combined value from both components. For the special case of the blend of Ly$\beta$ and the OVI doublet (J1001+5028 and J1031+3807), in order to calculate the REW for each ion, we drew a division at the velocity of minimum absorption between the two ions. The data listed under the heading Final Fit Parameters were derived from steps 3-4 above where information on $C_0<1$ is available, while the data under Initial Fit Parameters were derived from steps 1-2 where $C_0$ is unknown and $C_0=1$ is assumed. Note that the values of $\tau_0$ and $N_{\rm ion}$ derived from the $C_0=1$ fits are conservative lower limits. Also note that, the fits in Figure \ref{fig:fits} and their resulting parameters in Table \ref{table:lines} pertain only to the near-simultaneous, combined spectra described in \S\ref{data} (Figure \ref{fig:hst_mdm}). 

One basic result from this analysis is that \textit{OVI absorption is present and stronger than CIV in all six of the mini-BAL quasars.} In particular, the measured REW(OVI) values are 2-6 times larger than REW(CIV). This could be caused by larger covering fractions in OVI compared to CIV, larger column densities in OVI, or some combination of these effects. In three of the six mini-BALs (J0909+0002, J1001+5028, and J1031+3807), the OVI lines are resolved sufficiently to indicate saturation with $\sim$1:1 doublet ratios and \textit{partial covering}. The derived covering fractions in these cases are $C_0\sim0.4$ to 0.7. (The covering fractions in the other 3 mini-BAL quasars are undetermined.) Assuming the same covering fraction in all lines of a given system, the larger REWs in OVI in our fits correspond to 4-15 times larger column densities in OVI compared to CIV. 

The only other line securely identified in the mini-BAL quasars, besides CIV and OVI, is NV in J0909+0002 and J1031+3807. We also measure NV for J0906+0259, but it is tentative because it depends on an uncertain continuum (see Figures \ref{fig:hst_mdm} and \ref{fig:fits}). The Ly$\beta$ reported in J1001+5028 and especially in J0909+0002 is also very tentative because the measured features could be due entirely to unrelated Ly$\alpha$ forest lines (see Figure \ref{fig:fits}). 

Another {\it very} tentative mini-BAL detection is PV 1118,1128 in J1001+5028 (Figure \ref{fig:fits}). Our fit to this doublet (using CIV as a template with $C_0=1$, as shown in Figure \ref{fig:fits}) is also highly uncertain due to noise and contamination by the Ly$\alpha$ forest. The results should be regarded as an upper limit. There are no other candidate PV detections among the mini-BAL quasars. However, we point out that in the BAL quasar (J1301+0002, \S\ref{bal_fits}), the PV line is securely measured at a strength of only $\sim$1/3 of CIV (based on the line depths at the PV velocities). This amount of PV absorption relative to CIV could be present but undetected in all of the mini-BALs. 

\subsubsection{The BAL Quasar: J1301+0002} \label{bal_fits}
\begin{figure*}
	\includegraphics[width=1.01\linewidth, keepaspectratio]{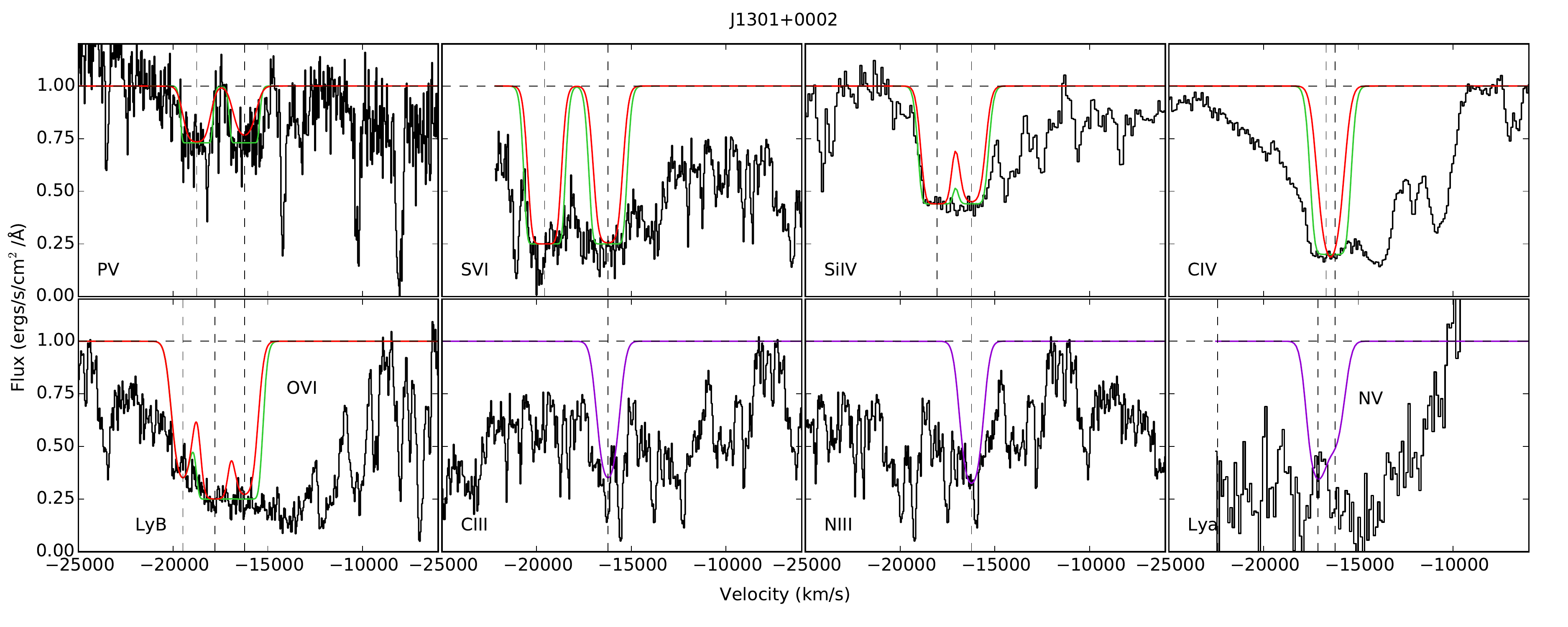}
	\caption{Fits to the absorption lines in the normalized combined spectra of the BAL quasar, J1301+0002. The flux is normalized by a continua described in \S \ref{measure}. The red curves are conservative lower-limit fits to the doublets where $b$ and the centroid velocity are determined from the PV lines while the covering fractions, $C_0$, and line-centre optical depths, $\tau_0$, are varied by eye so that the weaker (longer wavelength) components barely reach the data. These fits yield firm lower limits on $\tau_0$ and direct measurements of $C_0$ for each doublet. The green fits are illustrations of how larger $\tau_0$ values are consistent with and generally preferred by the data (see Table \ref{table:lines} for specific parameters used). The purple curves in the bottom row are fits to singlets and doublets that adopt $C_0$ from the OVI lines and only $\tau_0$ is varied (see \S\ref{bal_fits}).}
	\label{fig:bal_fits}
\end{figure*}

\begin{table*}
	\caption{Line Fit Parameters as described in \S\ref{mini_fits} and \S\ref{bal_fits}. Columns 1-4 are the basic fit parameters of the ``initial fit" where the shift is the redshift of the line in the quasars rest frame in km/s and $b$ is the Doppler parameter. Columns 5-9 are the ``initial fit" parameters which are generally $C_0$ = 1 where $C_0$ is the line of sight covering fraction, $\tau_0$ is the line centre optical depth, $N_{\rm ion}$ is the column density of that specific ion, and REW is the rest equivalent width derived from the fit. Note that $\tau_0$ and $N_{\rm ion}$ are lower limits. If the data provided enough constraints, columns 9-12 are the ``final fits" which have a $C_0$ < 1. If the doublet is resolved, $\tau_0$ is dictated by the longer wavelength component. The quasars with mini-BALs are above the double line and the BAL quasar is below the double line.}
	\label{table:lines}
	\begin{tabular}{*{12}{c}} 
	\toprule
	\multicolumn{4}{c}{Basic Parameters} & \multicolumn{4}{c}{Initial Fit Parameters} & \multicolumn{4}{c}{Final Fit Parameters} \\
	\cmidrule(lr){1-4} \cmidrule(lr){5-8} \cmidrule(lr){9-12}
	1 & 2 & 3 & 4 & 5 & 6 & 7 & 8 & 9 & 10 & 11 & 12 \\
	\midrule
	Quasar & Shift & $b$ & Ion & $C_0$ & $\tau_0$ & $N_{\rm ion}$ & REW & $C_0$ & $\tau_0$ & $N_{\rm ion}$ & REW \\
	 & (km/s) & (km/s) & & & & cm$^{-2}$ & ($\AA$) & & & cm$^{-2}$ & ($\AA$)\\
	\toprule
	\multirow{2}{*}{J0208+0022} & \multirow{2}{*}{-21905} &\multirow{2}{*}{2451} & CIV & 1.0 & 0.03 & 2.77E14 & 1.02 & - & - & - & -  \\
	 & & & OVI & 1.0 & 0.23 & 5.6E15 & 5.63 & - & - & - & -  \\
	 \midrule
	\multirow{3}{*}{J0906+0259} & \multirow{3}{*}{-20729} &\multirow{3}{*}{4172} & CIV & 1.0 & 0.02 & 3.78E14 & 1.4 & - & - & - & -  \\
	 & & & OVI & 1.0 & 0.12 & 4.88E15 & 5.34 & - & - & - & -  \\
	 & & & NV & 1.0 & 0.1 & 2.87E15 & 5.06 & - & - & - & -  \\
	\midrule
	\multirow{4}{*}{J0909+0002} & \multirow{4}{*}{-21724} &\multirow{4}{*}{1162} & CIV & 1.0 & 0.16 & 8.52E14 & - & 0.39 & 0.55 & 2.89E15 & 2.98  \\
	 & & & OVI & 1.0 & 0.2 & 2.27E15 & - & 0.39 & 5.42 & 6.14E16 & 6.52 \\
	 & & & NV & 1.0 & 0.16 & 1.28E15 & - & 0.39  & 0.7 & 5.61E15 & 3.7  \\
	 & & & Ly$\mathrm{\alpha}$ & 1.0 & 0.08 & 1.23E14 & - & 0.39 & 0.22 & 3.37E14 & 0.61 \\
	\midrule
	\multirow{2}{*}{J1000+1247} & \multirow{2}{*}{-7369} &\multirow{2}{*}{1359} & CIV & 1.0 & 0.05 & 3.02E14 & 1.1 & - & - & - & - \\
	 & & & OVI & 1.0 & 0.1 & 1.26E15 & 2.05 & - & - & - & - \\
	 \midrule
	\multirow{3}{*}{J1001+5028} & \multirow{3}{*}{-19661} &\multirow{3}{*}{729} & CIV & 1.0 & 0.15 & 5.06E14 & - & 0.67 & 0.25 & 8.25E14 & 1.87 \\
	 & & & OVI & 1.0 & 0.65 & 4.62E15 & - & 0.67 & 1.75 & 1.24E16 & 6.14 \\
	 & & & Ly$\mathrm{\beta}$ & 1.0 & 0.3 & 1.8E15 & - & 0.67 & 0.4 & 2.4E15 & 0.83 \\
	\midrule
	\multirow{3}{*}{J1031+3807} & \multirow{3}{*}{-22797} &\multirow{3}{*}{1039} & CIV & 1.0& 0.31 & 1.44E15 & - & 0.7 & 0.5 & 2.35E15 & 4.53 \\
	 & & & OVI & 1.0 & 0.43 & 4.26E15 & - & 0.7 & 0.95 & 9.63E15 & 6.56  \\
	 & & & NV & 1.0 & 0.3 & 2.15E15 & - & 0.7 & 0.4 & 2.86E15 & 4.27 \\
	\midrule
	\midrule
	\multirow{9}{*}{J1301+0002} & \multirow{9}{*}{-16229} &\multirow{9}{*}{596} & CIV & 1.0 & 1.0 & 2.7E15  & - & 0.8 & 7.0 & 1.89E16 & 9.19 \\
	 & & & OVI & 0.75 & 3.5 & 2.03E16 & - & 0.75 & 15.0 & 8.72E16 & 9.81 \\
	 & & & PV & 0.27 & 2.0 & 3.19E15 & - & 0.27 & 5.0 & 7.99E15 & 3.48 \\
	 & & & SiIV & 0.56 & 4.0 & 4.45E15 & - & 0.56 & 10.0 & 1.11E16 & 9.83  \\
	 & & & NV & 1.0 & 0.16 & 4.11E15 & - & 0.75 & 1.0 & 4.11E15 & 3.7 \\
	 & & & Ly$\mathrm{\beta}$ & - & - & - & - & 0.75  & 15.0 & 7.36E16 & 2.14 \\
	 & & & NIII & - & - & - & - & 0.75 & 2.3 & 7.52E15 & 2.95  \\
	 & & & CIII  & - & - & - & - & 0.75 & 2.0 & 1.07E15 & 2.72 \\
	 & & & SVI & 0.75 & 5.0 &  9.8E15 & - & 0.75 & 20.0 & 3.92E16 & 10.33 \\
	\bottomrule
	\end{tabular}
\end{table*}

The BAL quasar, J1301+0002, is a special case with many lines exhibiting a range of complex velocity profiles. These profiles and our fits to them (described below) are plotted in Figure \ref{fig:bal_fits}. Inspection of Figure \ref{fig:bal_fits} shows that the PV doublet is securely detected and resolved, revealing a $\sim$1:1 ratio that indicates saturation. This also implies that the lines of abundant ions like CIV and OVI have extreme optical depths, with $\tau_0 >$ several hundred, at least at velocities where PV is detected (\S\ref{intro}, \citealt{Hamann98}, \citealt{Leighly11}). The $\sim$1:1 doublet ratios indicating saturation are also evident in SVI and SiIV. Thus, we see that all of the measured BALs in J1301+0002 are optically thick at the PV velocities, and that their different depths (at these velocities) are due to different covering fractions. This has important consequences for the outflow structure, total column density, and kinetic energy. 

We adopt a fitting procedure based on the PV lines to focus on this high-column density portion of the BAL outflow. In particular, we use ``initial fits" to the PV doublet as a template for fits to the other BALs in limited velocity ranges that coincide with PV. The fitting steps are described below. 

\begin{enumerate}
	\item \textit{Initial fits:} First we fit the PV doublet to derive $b$, the velocity shift, and a lower limit on $\tau_0$ in the weaker 1128 \AA\ component. This PV fit is shown by the red curve in Figure \ref{fig:bal_fits}. It is achieved by letting $C_0$ be a free parameter, locking the doublet optical depths together in a 2:1 ratio, and then manually scaling $\tau_0$ so that the long-wavelength doublet component barely touches the data. We then apply $b$ and the velocity shift from this PV fit to CIV and the widely separated doublets SVI, OVI, and SiIV. We adopt the same strategy of letting the observed line depths determine $C_0$ and manually scaling $\tau_0$ to barely touch the data in the weaker doublet components. These fits are again shown by red curves in Figure \ref{fig:bal_fits}. They yield accurate values of $C_0$ but conservative lower limits on $\tau_0$ and the column density (columns 5-8 of Table \ref{table:lines}). 
	\item  \textit{Final fits:} We derive final fits to the BALs in two ways. For the lines with initial fits described above, we use the same $b$ value but manually increase $\tau_0$ so that the long-wavelength doublet components and overall doublet profiles have flatter bottoms (and more persistently deep troughs) that better match the data. These fits (shown by the green curves in Figure \ref{fig:bal_fits}) are only illustrative to show that larger values of $\tau_0$ (compared to the initial fits) are preferred by the data. They also provide better estimates of the REWs for just this high-column density portion of the outflow. However, the specific $\tau_0$ values adopted for these fits and tabulated in Table \ref{table:lines} are very poorly constrained by the data and are, in any case, still lower limits. The second type of final fit applies to CIII, NIII, and Ly$\alpha$, which are single lines poorly measured due to blends with other lines. For these BALs, we adopt the largest derived $C_0 =0.75$ (from OVI and SVI) and manually scale $\tau_0$ in the PV profile template to match the data. These fits (shown by the purple curves in Figure \ref{fig:bal_fits}) provide lower limits to $\tau_0$ and the ionic column densities. All of the final fit results are listed in columns 9-12 in Table \ref{table:lines}. 
\end{enumerate}

As with the quasars that contain the mini-BALs, OVI is stronger than CIV in REW but only by a factor of $\sim$1.1 for the final fits. Consequently, the column density of OVI is bigger than that of CIV by a factor of $\sim$5. For almost all ions, we found all lower limit optical depths to be $\tau_0$ > than 1.

The lower limit to the PV optical depth is an important parameter for our discussion of the outflow properties in \S\ref{disc}. Our initial fit yielded $b=596$ km/s and $\tau_0 = 2$ in the weaker doublet component, 1128 \AA\ (see the red curve in Figure \ref{fig:bal_fits}). However, this optical depth seems too small because the 1128 \AA\ doublet component in the initial fit is noticeably too weak. Our final fit shows that $\tau_0(1128) = 5$ provides a much better match to the data (green curve in Figure \ref{fig:bal_fits}). Due to this final fit and the fact that the data indicate a 1:1 doublet ratio which requires a $\tau_0 \ga$ 3, we adopt hereafter $\tau_0(1128) \ga$ 3 as a firm lower limit. Additional fitting experiments show that larger $\tau_0(1128)$ can also match the data if we allow smaller Doppler $b$ values. For example, optical depths up to $\sim$80 yield good fits (similar to the green curve) if $b$ is decreased to $\sim$400 km/s. We do not consider optical depths larger than 80 because $\tau_0$ > 80 corresponds to a total column density of roughly $N_H \ga 3\times 10^{23}$ and thus a Thompson scattering optical depth of $\ga$0.2 (based on photoionization analysis,  e.g., \citealt{{Leighly11}}, Hamann et al., in prep., see also \S\ref{structure}). These very large PV optical depths and column densities would place the absorber in a Compton thick regime that cannot produce absorption lines and are therefore ruled out. Thus, \textit{we conclude that the line-centre optical depth in PV 1128 is in the range $3\la \tau_0\la 80$. }

\subsection{CIV Variability} \label{variability}

We assess the variability in the CIV absorption lines by matching spectra from different epochs at wavelengths near (but outside) the CIV features. We adopt the well-calibrated SDSS spectrum of each quasar as the flux standard. Other spectra are scaled to match this standard by fitting power laws constrained by the median flux in three narrow spectral windows that appear free of strong emission and absorption (with widths of $\sim$25-50 \AA\ centred at $\sim$1350 $\AA$,  $\sim$1450 $\AA$, and $\sim$1600 $\AA$ in the quasar rest frame). This procedure usually produces good matches for our variability analysis. However, in some cases, we make additional adjustments by vertically scaling the spectra (J0208+0022) or by slightly shifting the windows used to anchor the power law fits (e.g., if the windows appear contaminated by emission or other absorption lines that also might have varied) in order to showcase the CIV features.

Since the SDSS epoch is not contemporaneous with the HST observations, there can be mismatches between the HST and the contemporaneous ground-based spectra (e.g., J0909+0002 -- see Figure \ref{fig:hst_mdm}). This could be a real variation in the flux, but we do not adjust for this as it does not affect the analysis carried out in this paper. We do however scale the ground-based spectra to the SDSS spectrum for each quasar as we are interested in the relative flux variability in this paper. 

\begin{figure*}
\begin{center}
 \includegraphics[scale=0.4,angle=-0.0]{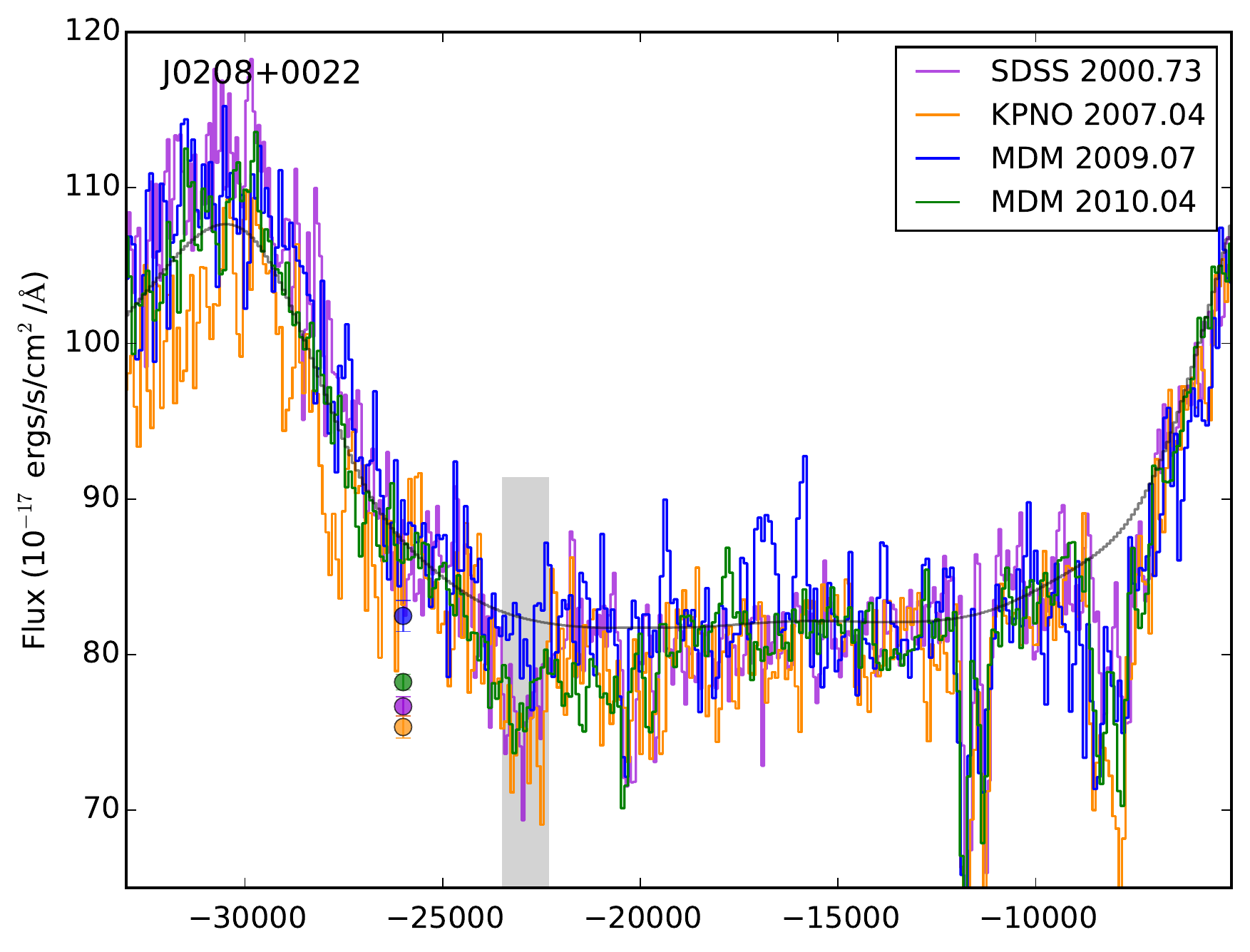}
 \includegraphics[scale=0.4,angle=-0.0]{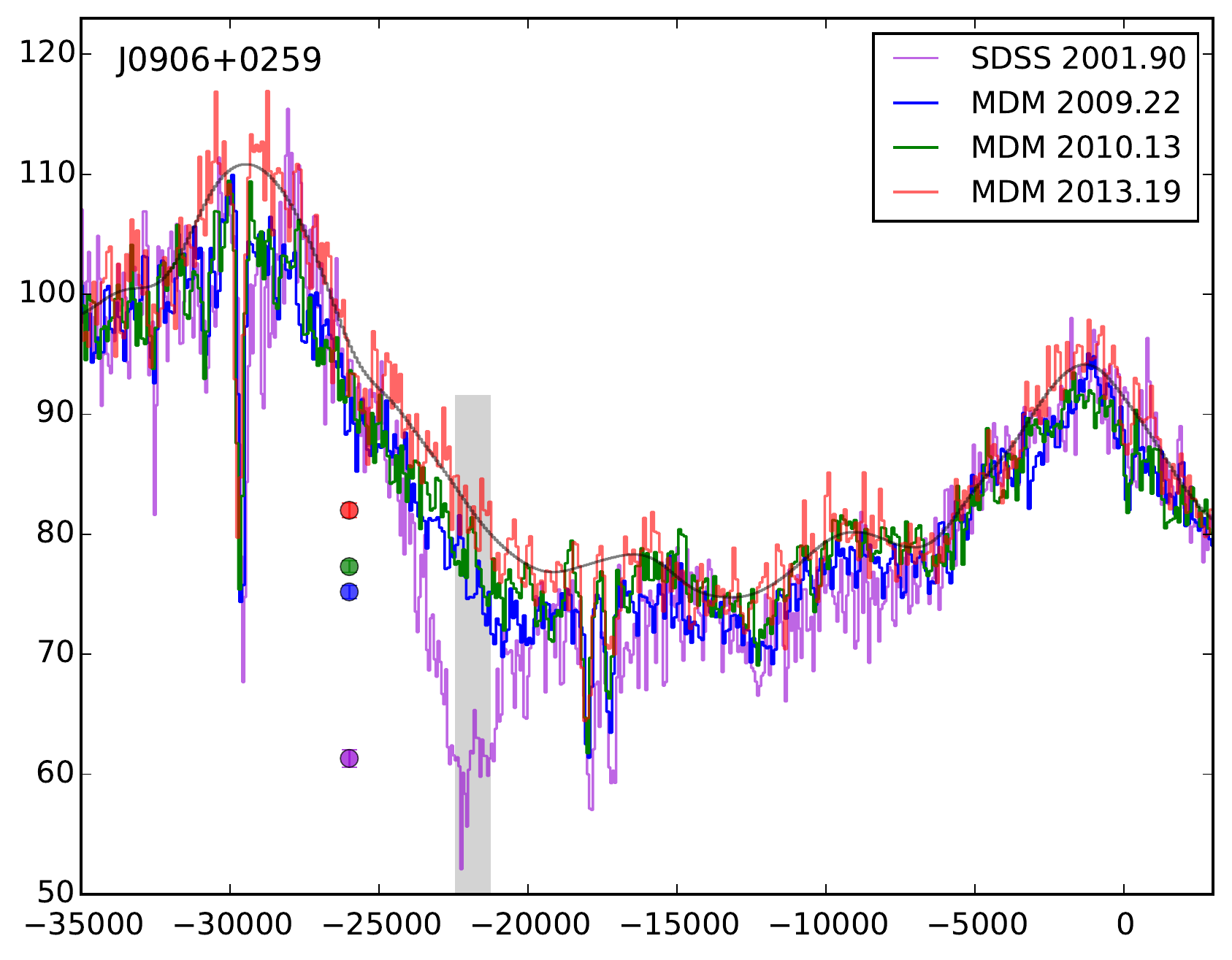}
 \includegraphics[scale=0.4,angle=-0.0]{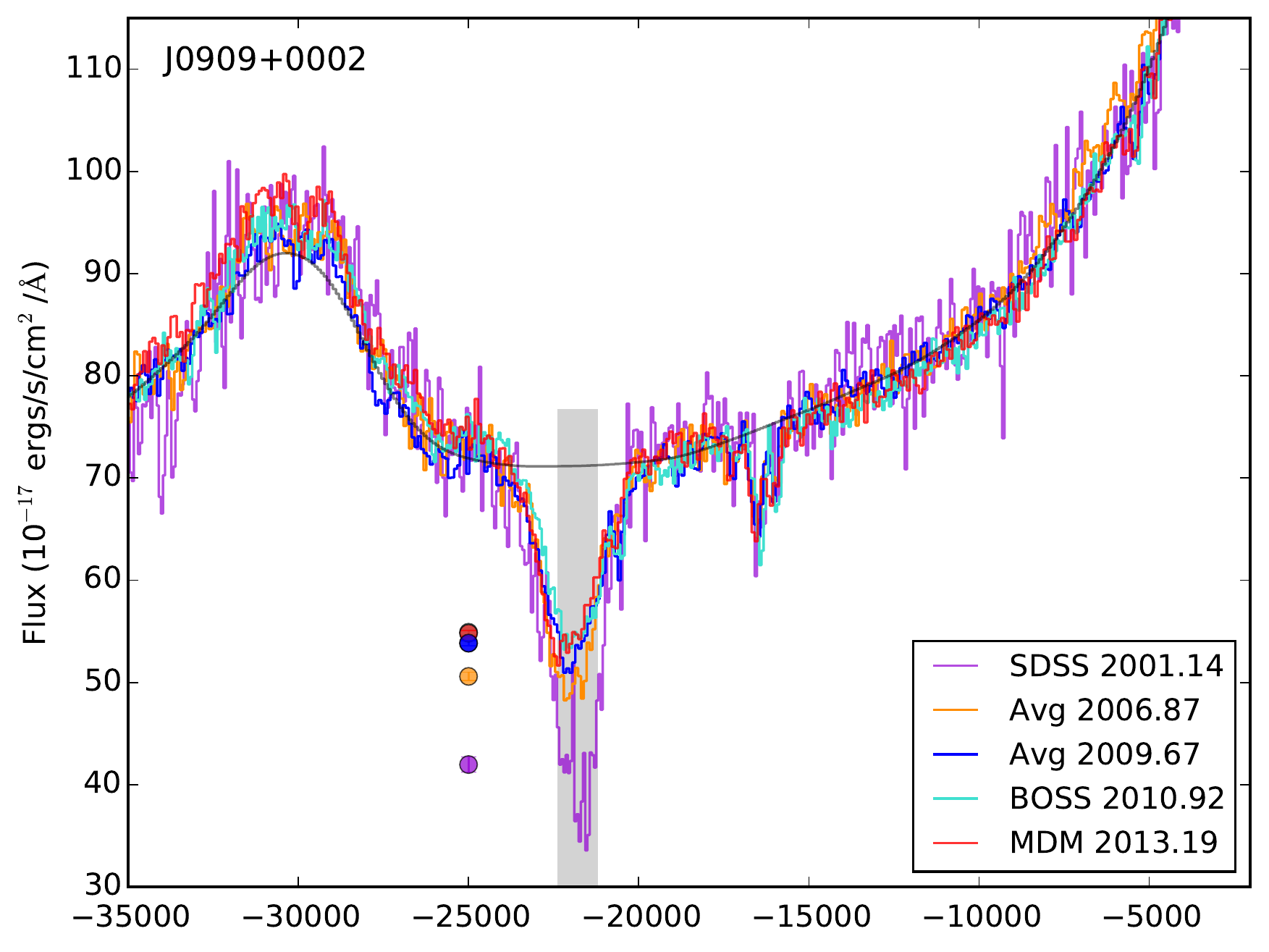}
 \includegraphics[scale=0.4,angle=-0.0]{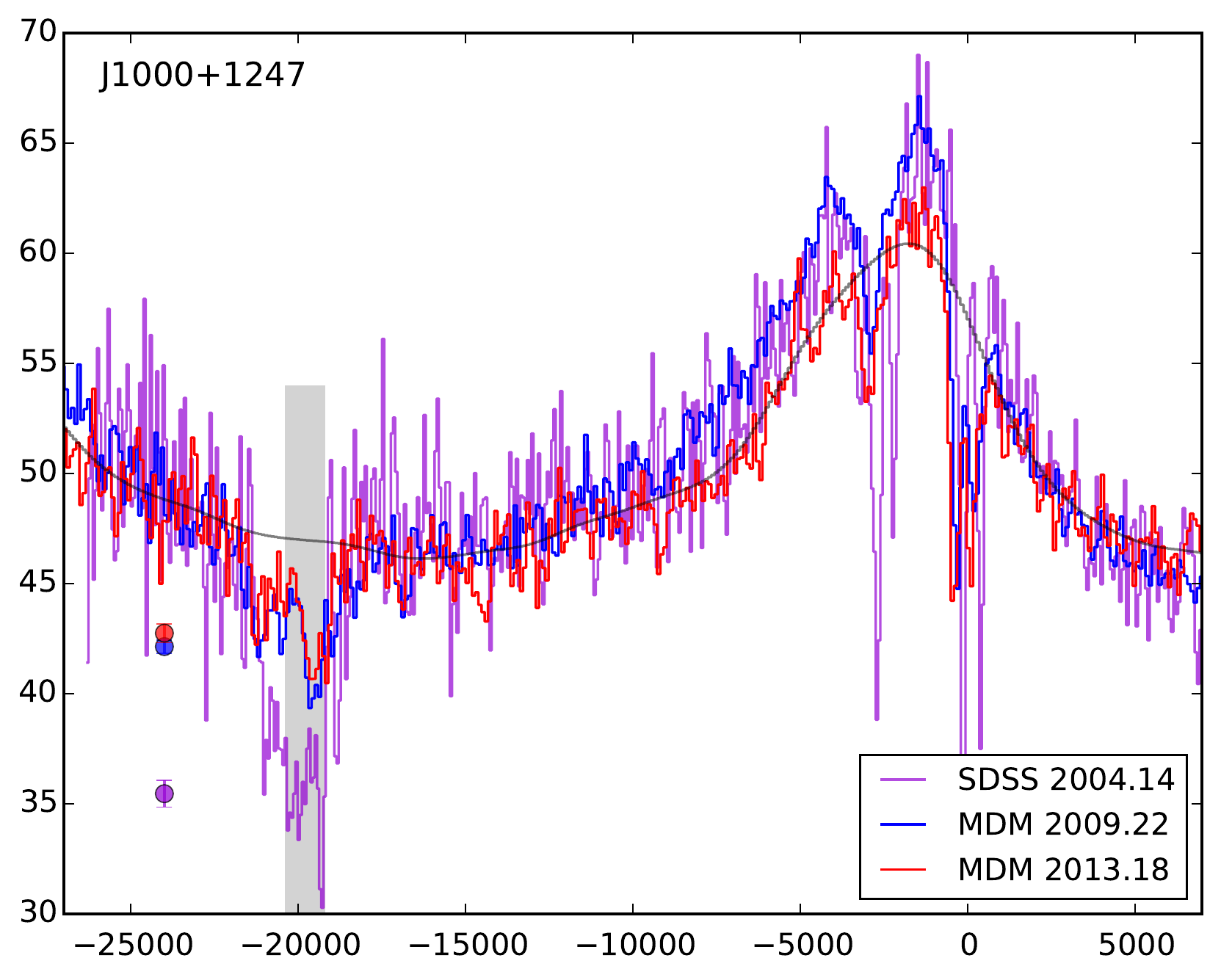}
 \includegraphics[scale=0.4,angle=-0.0]{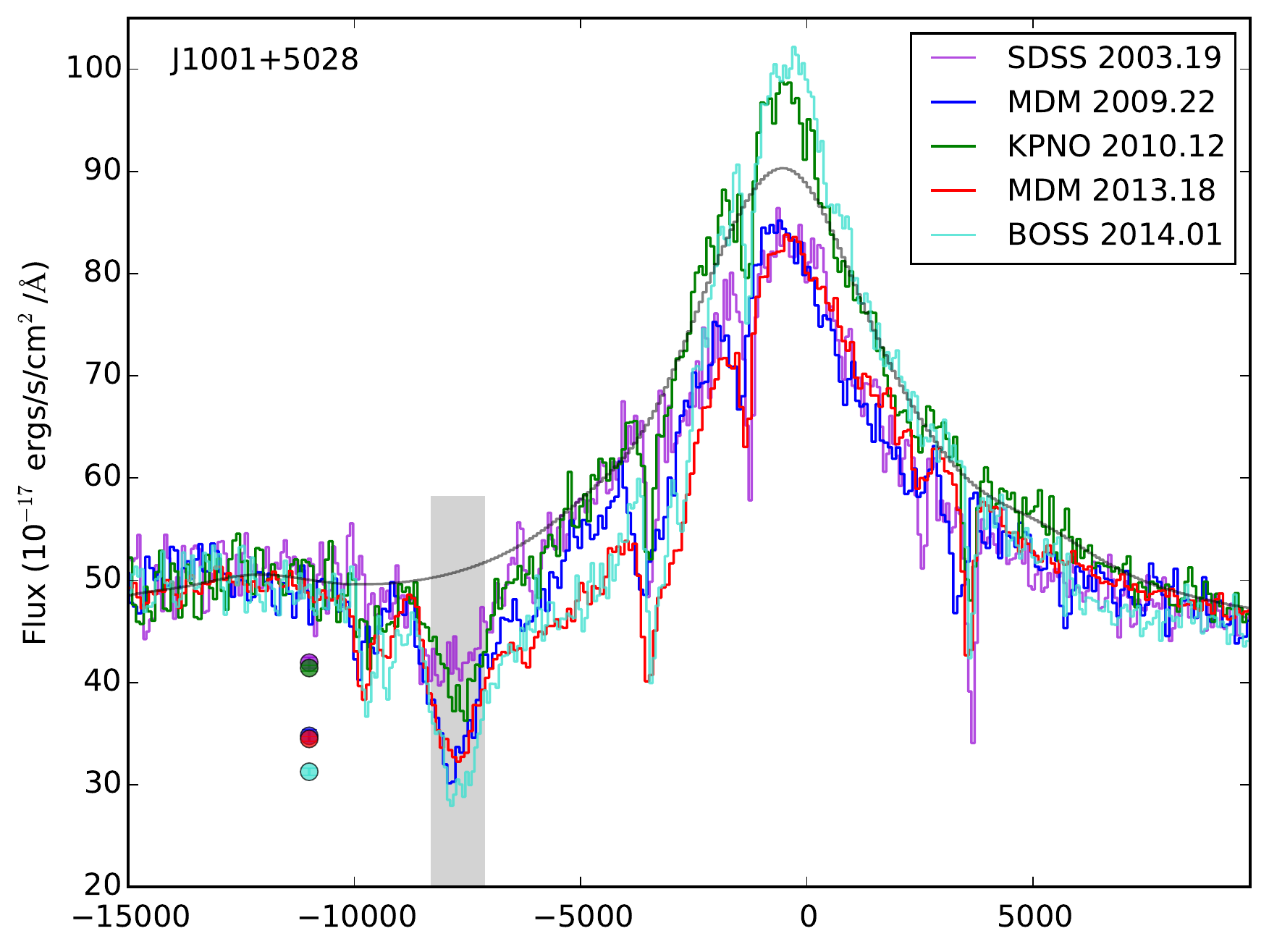}
 \includegraphics[scale=0.4,angle=-0.0]{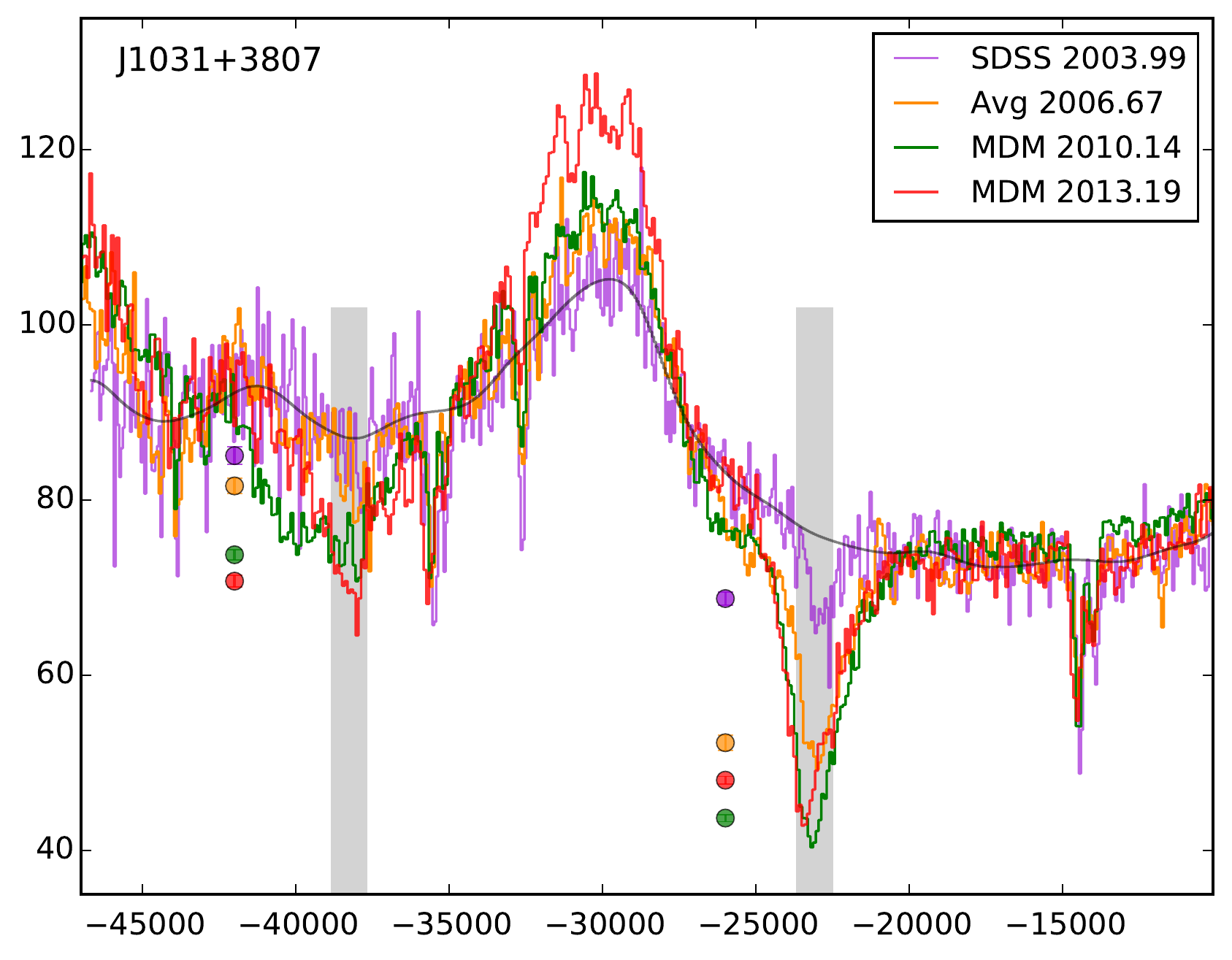}
 \includegraphics[scale=0.4,angle=-0.0]{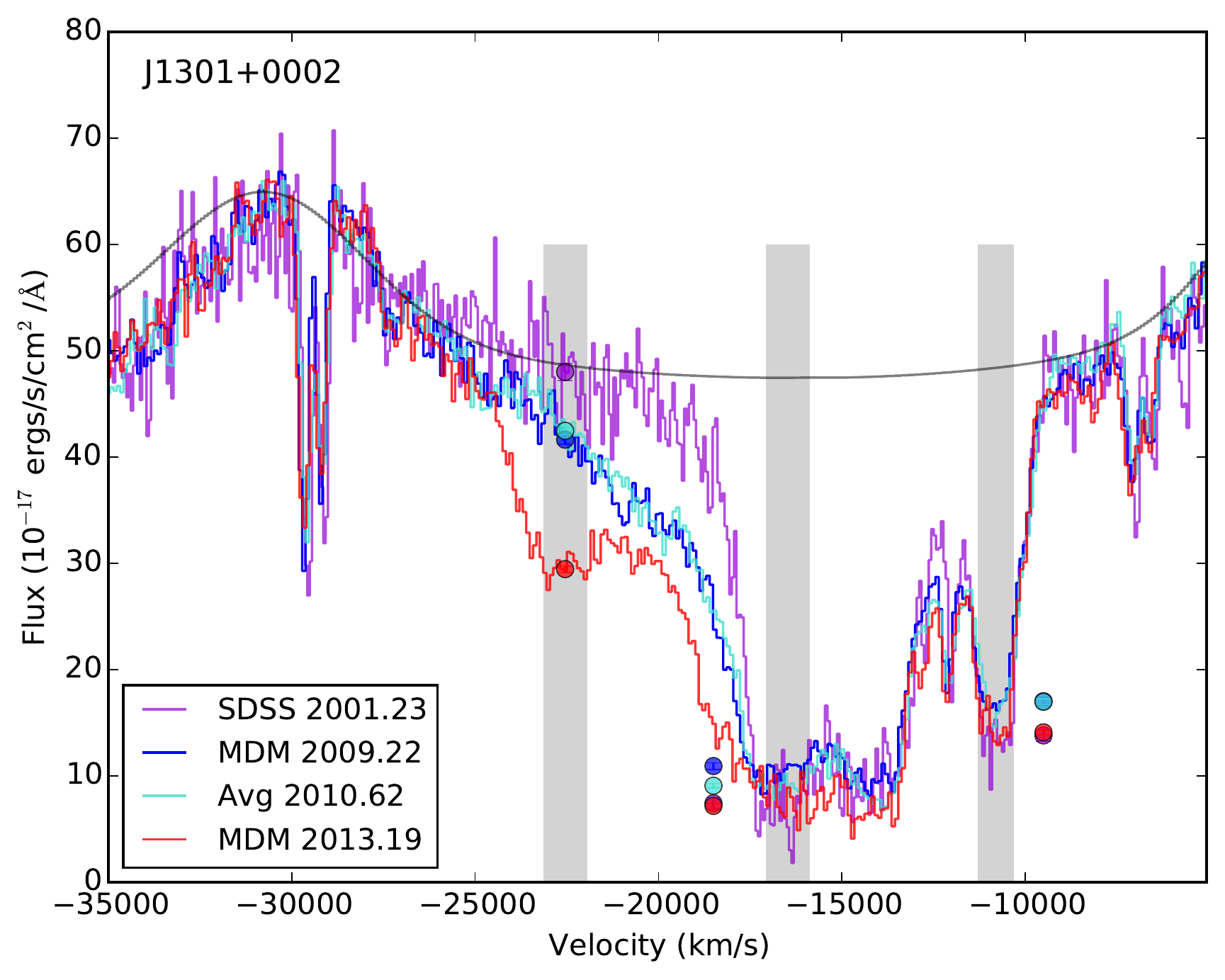}
 \end{center}
 \vspace{-12pt}
 \caption{The various epochs of each quasar showing the variability of CIV.  The solid black lines show the continua used to measure $\Delta A$ and are the same continua show in Figure \ref{fig:hst_mdm}, except for J1031+3807 (see \S\ref{variability} for details). The circular dots are the median flux within the variability window of $\sim$1200 km/s. The variability windows are marked by the grey shaded regions. The velocity is with respect to the CIV longer wavelength doublet line and each quasar's redshift. Error bars are plotted, but in general are small. Every single targeted CIV feature varied. There is an emergence of a ``high-velocity" BAL in J1031+3807.}
 \label{fig:avg_flux}
\end{figure*}

Figure \ref{fig:avg_flux} shows the matched spectra around the broad CIV absorption lines in all seven quasars. Inspection of this figure indicates that CIV variability occurred in every quasar. In particular, three outflows exhibited a strengthening in CIV (J1001+5028, J1031+3807, J1301+0002), three became weaker (J0906+0259, J0909+0001, J1000+1247), and one varied, but returned approximately to its original strength (J0208+0022). In the BAL quasar, J1301+0002, there are several components of CIV that vary and the middle variable component (-17000 km/s) corresponds to the same redshift as that of the PV absorption.

\begin{figure}
 \includegraphics[width=\linewidth, height= \linewidth, keepaspectratio]{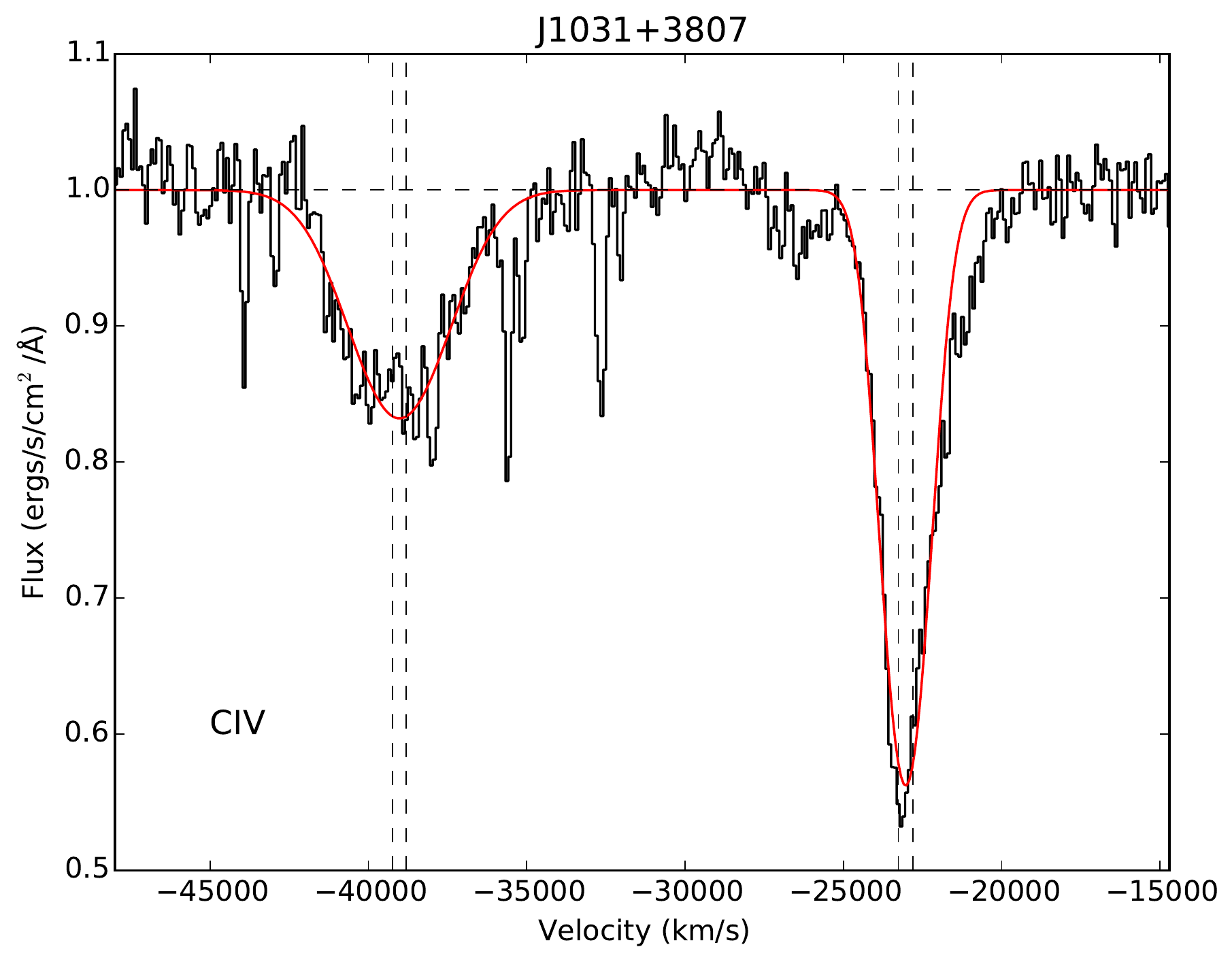}
 \caption{MDM 2010.14 epoch used to showcase the high-velocity BAL that emerged in J1031+3807. The redward feature is the mini-BAL and the blueward feature is the high velocity BAL. The flux is normalized by the continuum shown in Figure \ref{fig:hst_mdm} and described in \S\ref{mini_fits}. The red curve is the fit to the CIV mini-BAL and high-velocity BAL used to measure the REW and FWHM of the high-velocity BAL (see \S\ref{variability}). The dotted lines are the resultant centroids from fits to the two CIV features. }
 \label{fig:103112_double_c4}
\end{figure}

In the mini-BAL quasar, J1031+3807, a new, broad, and extremely blueshifted CIV outflow line appeared at the velocity shift of v~$\approx -38800$ km s$^{-1}$ with a FWHM $\approx 3830$ km s$^{-1}$. The variability in this new high velocity BAL roughly followed that of the CIV mini-BAL, accentuated by the fact that the high-velocity BAL was strongest in our MDM 2010.14 spectrum when the mini-BAL at v~$\approx -22800$ km s$^{-1}$ was also strongest (see Figure \ref{fig:103112_double_c4}). Figure \ref{fig:103112_double_c4} shows fits to both of the broad outflow features in the MDM 2010.14 spectrum after normalizing by the continuum plotted in Figure \ref{fig:avg_flux}. The fit to the CIV mini-BAL is the same fit shown in Figure \ref{fig:fits}. The fit to the high-velocity BAL is follows the same fitting strategy as that in \S\ref{mini_fits} with two Gaussians separated by the CIV doublet separation. We identify the new BAL as high-velocity CIV instead of moderate-velocity SiIV (crudely 18,000 km/s) because SiIV absorption should be accompanied by even stronger CIV at the same velocities, which is not observed (\citealt{Hamann08}, \citealt{Hamann13}). 

We quantify the line variabilities by measuring the flux in spectral windows shown by the grey vertical bars in Figure \ref{fig:avg_flux}. These spectral windows are centred on the deepest average absorption of the observed troughs within a window of $\sim$1200 km/s in the epochs when the troughs were strongest. The velocity width of $\sim$1200 km/s is at least three times larger than the spectral resolution and is similar to the minimum width over which large variations occurred in our data (see also \citealt{Capellupo11}). In order to obtain good flux measurements with small uncertainties, the width of $\sim$ 1200 km/s was chosen as a compromise such that it is narrow enough to be representative of the variability, yet not too wide such that the dynamics of the variability get smoothed out. The exact widths of these windows are not important for our analysis or the results. 

There was one exception to this spectral window width rule, J0208+0022. Even though MDM 2010.04 has the deepest median absorption, it has a noisy absorption profile. Therefore, we chose the SDSS epoch to define the variability window as it has the most well defined feature and is roughly the same average depth as that of MDM 2010.04. J1031+3807 has two such flux windows to characterize the two distinct absorption troughs. The BAL quasar, J1301+0002, shows various amounts of variability at different velocities in the broad CIV BAL. Therefore in this quasar, we define three flux windows that roughly align with absorption minima (Figure \ref{fig:avg_flux}). The middle window covers the velocities with PV absorption (Figure \ref{fig:fits}). The lowest-velocity window in J1301+0002, at roughly $-$10800 km s$^{-1}$, is only 1000 km s$^{-1}$ wide because that absorption feature is narrow.  

The coloured dots next to each outflow line in Figure \ref{fig:avg_flux} indicate the median fluxes measured at different epochs in the grey windows. These flux points are plotted with 1$\sigma$ error bars too small to be visible in the figure. The errors are based only on the photon statistics,
\begin{equation}
	\sigma^2 = \frac{1}{n^2} \sum\limits_{i=1}^n \sigma_i^2
\end{equation}
where $\sigma_i$ is the error at pixel $i$ and $n$ is the number of pixels included in the flux window. For the KPNO spectra, which do not have error spectra available, the errors are estimated from pixel-to-pixel fluctuations in the flux.

We convert the median fluxes to normalized absorption strengths, $A$, that represent the depth of the line below the continuum in the normalized spectra (with values $0 \lesssim A \lesssim 1$). The continua used for the normalization are shown by the smooth black curves in Figure \ref{fig:avg_flux} and are the same continua described in \S\ref{measure}, except in the case of  J1031+3807 where a different continuum had to be determined due to changes in the line profile throughout time at approximately $-$25,000 km/s. We calculate the change in depth, $\Delta A$, and its associated error between all observed consecutive epochs using these continua. Following the procedure in \cite{Capellupo11}, we consider line variations between two epochs to be real (or ``secure") if $\Delta A$ is larger than its $4\sigma$ error {\it and} they pass a visual inspection to rule out anomalies such as poor spectral matching or variable broad emission lines at the absorption line wavelengths. All variability discussed hereafter meets this $4\sigma$ criteria. 

Table \ref{table:quasar_observations} summarizes some of the variability results for each quasar. In particular, $\Delta t_{\rm shortest}$ and $\Delta A_{\rm shortest}$ are the time difference and normalized amplitude change corresponding to the shortest time scale of secure variations, and $\Delta t_{\rm largest}$ and $\Delta A_{\rm largest}$ are the time and amplitude differences corresponding to the largest amplitude change detected in that quasar. The observing epochs used for these measurements are listed in parentheses next to each $\Delta t$. $\Delta t_{\rm total}$ is not a variability time, but rather the total time between our first and last observation of that quasar. All of the time differences listed here and discussed in the remainder of this paper are in the quasar rest frame. Some of the variability measurements involve average spectra constructed from two observations on different dates ({\S\ref{data}). In these case, we derive $\Delta t$ from the observation dates that were closest in time between the two epochs. 

In the mini-BAL quasars, we find secure 4$\sigma$ variations across times as short as $\Delta t_{\mathrm{shortest}}$ = 0.29 yr with an amplitude change of $\Delta A_{\mathrm{shortest}}$ = 0.063 $\pm$ 0.008 (in J1001+5028), and find amplitude changes as large as $\Delta A_{\mathrm{largest}}$ = 0.33 $\pm$ 0.01 on a time scale of $\Delta t_{\mathrm{largest}}$ = 2.12 yrs (in J1031+3807). The largest amplitude variation overall in our data occurred in the high-velocity window of the BAL quasar, J1301+0002, with $\Delta A_{\rm largest}$ = 0.38 $\pm$ 0.018 in time $\Delta t_{\mathrm{largest}}$ = 4.29 yrs. The shortest time scale variation listed for the BAL quasar, with $\Delta t_{\mathrm{shortest}}$ = 0.33 yr and $\Delta A_{\mathrm{shortest}}$ = 0.039 $\pm$ 0.008, occurred in the middle spectral window coincident with PV absorption. 

Every quasar in our sample exhibited secure CIV absorption line variability on a timescale of $\lesssim$ 1.9 years. All of the variations are consistent with a strengthening or weakening of the absorption troughs at a fixed velocity. Thus we find no evidence for acceleration or deceleration in the outflows. However, we do find complex variabilities where the shape of the absorption profile changed or absorption appeared or disappeared entirely at some velocities (see Figure \ref{fig:avg_flux}). For example, the mini-BAL in J1000+1247 grew narrower and its centroid shifted to lower velocities across our monitoring period. In J0906+0259, dramatic changes in both strength and profile of the mini-BAL were accompanied by amorphous low-level flux changes across a wide velocity range from at least $-$7500 to $-$30,000 km/s. Notice in particular that the CIV absorption minimum has shifted from $-$22,000 km/s in SDSS 2001.90 to $-$20,700 km/s in MDM 2009.22 and another minimum appeared at $\sim$12,200 km/s. The low-level flux changes in this quasar appear across a wide velocity range, are only on the blue side of the CIV emission line, and are not consistent with variations in weak emission lines at these wavelengths. Thus, we attribute all of the small flux difference at these velocities to variable CIV absorption. We also see this low level flux variation on the blue side of the CIV emission line in J0208+0022 from $-$23,000 to $-$19,000 km/s. Similar complex variations have been reported previously in BAL and mini-BAL quasars (e.g., \citealt{Paola11}, \citealt{Hall11}, \citealt{FilizAk12}, \citealt{FilizAk13}, \citealt{Hamann13}, \citealt{Rogerson16}). 

In J1031+3807, the mini-BAL near $-$22,800 km/s became dramatically stronger and broader while a new high-velocity BAL appeared at $-$38,800 km/s (see Figures \ref{fig:hst_mdm} and \ref{fig:103112_double_c4}). The mini-BAL grew, specifically, from REW $\approx 0.96$ \AA\ and FWHM $\approx 986$ km/s in the SDSS spectrum (\citealt{Paola09}, Table \ref{table:quasar_properties}) to 4.5 \AA\ and 1730 km/s, respectively (Table \ref{table:lines}), in our MDM observation $\sim$2.1 years later. The newly emerging CIV BAL in the MDM spectrum had REW $\approx 3.5$ \AA\ and FWHM $\approx 3830$ km/s. 

Our results overall are consistent with previous variability studies. For example, \cite{Misawa14} found that all of the 7 mini-BALs in their sample varied on a timescale of 1 to 3.5 years in the quasar rest frame. In an unusual sample of extreme high-velocity mini-BALs, \cite{Hamann13} reported that at least 5 of 7 quasars varied between two epochs roughly 2-3 years apart. \cite{Paola09} found that only 52\% of mini-BALs and borderline mini-BAL/BALs in a sample of 26 quasars varied between two observations across 1-3 year time scales. BAL variability has been studied more extensively, with reported variability fractions ranging from $\sim$92\% \citep{Gibson08} to 67\% \citep{Capellupo11} to 50-60\% (\citealt{FilizAk13}) on time scales up to a few years. 

The differences between these studies can be attributed to stochastic behaviours in small samples and to known trends for greater variability in outflow lines that are weaker and/or at higher velocities. For example, \cite{Capellupo11} and \cite{Capellupo13} report that the variability rates depend on both monitoring time scale and the initial line strength (depth). In particular, they find variability fractions ranging from $\sim$35\% for time scales of $\Delta t$ = 0.3 yr, to $\sim$60\% for $\Delta t =1$ yr,  to $\sim$90\% for $\Delta t \sim 6$ yrs. Across all time scales $\lesssim$ 10 yrs, lines or portions of BAL troughs with average strengths A $\sim$ 0.2 are 4-5 times more likely to vary than deeper features with average A $\sim$ 0.8.

In our study, all of the mini-BALs varied, but half of them exhibited variability within a time of $\Delta t$ $\sim$ 1.1 years. The median amplitude change of the secure detections for the mini-BALs was $\Delta A\sim 0.12$. These results are approximate due to our small sample size and limited numbers of epochs, but they are consistent with the studies mentioned above. Thus we adopt $\Delta t \sim 1.1$ as ``typical'' variability time and $\Delta A\sim 0.12$ as a corresponding ``typical'' amplitude change for our discussions in \S\ref{disc} below. 

\section{Discussion} \label{disc}

\subsection{Implications for Outflow Structure}\label{structure}

The results in \S\ref{results} provide important information on the structure and physical conditions in the quasar outflows. One result is that OVI absorption is present and stronger than CIV in all six mini-BAL quasars, while lower ionization lines such as SiIV 1393,1403 and CII 1335 are not detected (\S\ref{mini_fits}, Figures \ref{fig:hst_mdm} and \ref{fig:fits}). High degrees of ionization are clearly favored. This preference for higher degrees of ionization and OVI being stronger than CIV was also seen in the study of a mini-BAL system in \cite{Paola11}. Our study reinforces this result by, for the first time, not only adding detections of OVI in mini-BALs systems, but also finding for the first time that OVI is generally stronger than CIV. Further, in our sample the OVI lines also appear saturated based on $\sim$1:1 doublet ratios in all three mini-BAL quasars where the doublet ratio is constrained. Saturation in shallow absorption troughs indicates that the absorber only partially covers the background light source. The derived OVI covering fractions are in the range 0.4 to 0.7 (Table \ref{table:lines}). 

In the BAL quasar, J1301+0002, more lines are detected including PV, SVI, and SiIV (Figures \ref{fig:hst_mdm} and \ref{fig:bal_fits}). All of these lines are optically thick, at least across the velocities where PV is present (\S\ref{bal_fits}). This is again evident from $\sim$1:1 doublet ratios, and from PV absorption whose low abundance signals saturation in the lines of abundant ions like CIV and OVI (see refs. in \S\ref{intro}). The range of observed depths in these BALs indicates a range of covering fractions, from 0.27 in the PV gas to 0.8 in CIV, in a spatially inhomogeneous outflow.

Inspection of Figure \ref{fig:hst_mdm} shows that all of these instances of partial covering pertain to the quasar continuum source and not the broad emission-line region (because there are no strong emission lines at the absorption line wavelengths). This requires small outflow absorbing structures because the UV continuum source (e.g., the thermally-emitting inner accretion disk) is expected to be only $<$ 0.01 pc across \citep[in luminous quasars like our sample, see][Hamann et al., in prep., and refs. therein]{Hamann11}. The range of covering fractions evident in the BAL quasar J1301+0002 requires inhomogeneities (i.e., substantial column density variations) on spatial scales $\lesssim$ 0.01 pc. The outflow structures at any given velocity might resemble the schematic illustrations in \cite{Hamann01} and \cite{Hamann04}, which depict inhomogeneous partial covering due to small outflow clumps or filaments. 

Regardless of the specific geometry, the BALs in J1301+0002 reveal that the highest column density regions traced by PV are the most spatially compact across a narrow range of velocities. The velocity dispersion in this gas, based on the measured PV line widths, has a FWHM of only $\sim$1300 km/s. In contrast, the low-column density gas traced by strong transitions like CIV and OVI occupies both a larger spatial area (leading to larger covering fractions and deeper absorption troughs) and a wider range of velocities (leading to broader profiles, e.g., FWHM~$\sim$ 9000 km/s in CIV). 

We can estimate the total column density, $N_H$, in the PV absorbing regions from photoionization models that assume solar abundances and a standard quasar spectral shape across the UV (\citealt{Hamann98}, \citealt{Leighly11}). In particular, Figure 15 in \cite{Leighly11} indicates that our estimate of $\tau_0(1128{\rm \AA}) \gtrsim 3$ in the weaker PV line requires $N_H\gtrsim 2.5\times 10^{22}$ cm$^{-2}$ (after adjusting for the different line width and combined doublet oscillator strength used by Leighly et al. in their calculations). This {\it minimum} total column density is similar to previous estimates based on PV measurements in BAL and mini-BAL outflows (e.g., \citealt{Hamann98}, \citealt{Hamann03}, \citealt{Leighly09}, \citealt{Leighly11}, \citealt{Borguet12}, \citealt{Capellupo14}, Hamann et al., in prep.). 

It is interesting to note that the PV lines in J1301+0002 would, by themselves, be classified as mini-BALs at v~$\sim 16,230$ km/s if they were not accompanied by broad BALs of CIV and other lines in the same spectrum. This mixture of line widths in one outflow supports the idea that BALs and mini-BALs in other quasars probe the same general outflow phenomenon. The main observational differences, namely narrower and often weaker troughs in the mini-BALs, might be caused either by smaller covering fractions or lower total column densities that produce mini-BALs instead of BALs. For example, we can imagine scaling down all of the column densities in the BAL outflow in J1301+0002 to produce a typical mini-BAL system. In particular, this scaling would result in the spatially-extended low-column density regions that now produce strong OVI and CIV BALs yielding no significant absorption at all, while the highest-column densities regions that now produce weak and narrow PV in J1301+0002 would instead yield weak and narrow mini-BALs in strong transitions of CIV and OVI. 

Alternatively, mini-BAL outflows might have generally similar column densities to BALs but with smaller outflow structures leading to smaller covering fractions and weaker and narrower absorption line profiles. This interpretation is supported by detections of PV absorption (and thus large column densities) in some individual mini-BAL outflows (Hamann et al., in prep.) and by the results in Herbst et al. (in prep.) showing that, in composite quasar spectra from the BOSS survey, PV absorption is often present (with $\sim$1:1 doublet ratios) across a wide range of outflow types from mini-BALs to the strongest BALs. 

In either case, mini-BALs could represent individual clumps or filaments in quasar outflows while BALs form along lines of sight that intercept larger clumps/filaments or perhaps larger ensembles of these structures (\citealt{Hall07}, \citealt{Hamann08}, \citealt{Hamann13}). If the differences between BALs and mini-BALs are caused by viewing angle effects, mini-BALs might form along the ragged edges of BAL outflows (e.g., at higher latitudes above the accretion disk) where there are smaller and/or fewer clumps or filaments along each line of sight (see also \citealt{Proga00}, \citealt{Ganguly01}, \citealt{Proga04}, \citealt{Chartas09}, \citealt{Hamann11}, \citealt{Hamann12}). 

The observed line variabilities provide further evidence for a close physical relationship between BALs and mini-BALs. This is most notable in extreme cases where one outflow type morphs into the other or where simple single mini-BALs are replaced by complex amorphous absorption across a range of velocities (e.g., \citealt{Leighly09}, \citealt{Hall11}, \citealt{Paola11}, \citealt{Paola13}, \citealt{Rogerson16} and refs. therein). In our data, we find one mini-BAL that developed weak BAL-like absorption across $\gtrsim$ 20,000 km/s (J0906+0259, Figure \ref{fig:avg_flux}) and another that grew dramatically stronger at v~$\approx -22,800$ km/s while a distinct new high-velocity BAL emerged at v~$\approx -38,800$ km/s in the same spectrum (J1031+3807). The strengthening of the mini-BAL coupled with the emergence of the high-velocity BAL in this latter case suggests that the two systems are somehow physically related even though they are distinct and well separated in velocity. This interpretation depends on the cause of the line variations (\S1 and \S\ref{locations}). However it is, in any case, another example of the complex and highly structured nature of these BAL/mini-BAL outflows.

\subsection{Variability and Absorber Locations}\label{locations}

\begin{table*}
	\caption{Outflow locations and crossing speeds for representative rest-frame timescales, $\Delta t$, and normalized line amplitude changes, $\Delta A$, in our data based on the `crossing disc' model for the line variations. See \S\ref{locations}.}
	\label{table:locations}
	\begin{tabular}{*{5}{c}} 
	\hline
	Case & $\Delta t$ & $\Delta A$ & Crossing speed & $\sim$ Location \\
	 & (yrs) & & (km/s) & (pc) \\
	\hline
	``Typical" mini-BAL & 1.1 & 0.12 & 3500 & 2.5 \\
	Shortest $\Delta t$ in mini-BALs (J1001+5028) & 0.29 & 0.063 $\pm$ 0.008 & 7640 & 0.4 \\
	Largest $\Delta A$ in mini-BALs (J1031+3807) & 0.81 & 0.22 $\pm$ 0.015 &3750 & 0.7 \\
	Shortest $\Delta t$ in BAL & 0.33 & 0.039 $\pm$ 0.008 & 4690 & 0.8 \\
	Largest $\Delta A$ in BAL & 4.29 & 0.38 $\pm$ 0.018 & 1110 & 14 \\
	\hline
	\end{tabular}
\end{table*}

Locations of the outflow gas derived from line variabilities depend on the cause of the variations. Two possible causes are outflow structures moving across the line of sight and ionization changes caused by changes in the incident ionizing flux (see \S\ref{intro}, also \citealt{Capellupo13}, \citealt{Capellupo14}, \citealt{Rogerson16}, McGraw et al., in prep.). Short-term variability caused by moving clouds indicates large crossing speeds close to the quasars; the derived distances for luminous quasars with variability times of order 1 yr are of order a few pc \citep[see below, also][]{Capellupo13}. 

Variations caused by ionization changes can occur at any distance within the quasar's radiative sphere of influence. In this case, the variability times place generally weak constraints on the gas densities (from the recombination time) leading to large upper limits on the distances. For example, for the quasars in our study with typical luminosities $\lambda L_{\lambda}\sim 5\times 10^{46}$ ergs/s (Table \ref{table:quasar_properties}) and typical rest-frame variability times $\sim$1.1 yr (\S\ref{variability}), the inferred densities are $n_e \ga 10^4$ cm$^{-3}$ (\citealt{Hamann95}, \citealt{Kriss95}, \citealt{Hamann97}) leading to distance upper limits\footnote{This result assumes the clouds are optically thin in the Lyman continuum and are photoionized by a standard quasar spectrum to an ionization level (described by ionization parameter $\log U \approx -1.6$) that is roughly optimal for CIV (see Appendix A in \citealt{Hamann11}). If the clouds are not optically thin, as expected for the high column densities in the BAL quasar J1301+0002 (\S\ref{structure}, also \citealt{Leighly11}, Hamann et al., in prep.), or they experience some other form of radiative shielding consistent with the general X-ray weakness of BAL quasars (\citealt{Murray95}, \citealt{Gallagher02}, \citealt{Gallagher06}), then much smaller distances could be derived.} $\lesssim$ 1 kpc. These results are marginally consistent with studies that use excited-state absorption lines to infer absorber distances, e.g., \cite{Moe09} and \cite{Dunn10} who derive $R\sim 3.3$ and $\sim$6 kpc, respectively. The most interesting comparisons would involve cases with short-time variability (such as J1001+5028 and J1301+0002 in Table 3) where the maximum distances inferred from ionization changes would be $\sim$300 pc. However, there are significant assumptions and uncertainties in these estimates and detailed comparisons to other work are beyond the scope of this study. For our current discussion, it is sufficient to note that $R \la 1$ kpc based on a typical time scale is an approximate typical upper limit to the absorber distances. 

Crossing clumps or streams provide a natural explanation for cases of dramatic variability, such as the emergence of a new high-velocity BAL in the mini-BAL quasar J1031+3807 (see also \citealt{Hamann08}, \citealt{Leighly09}, \citealt{Rogerson16}). They are are also favored by observations of variability in saturated lines that should be unresponsive to modest changes in their optical depths caused by modest ionization changes. However, with the exception of the BAL quasar J1301+0002, the optical depths in the variable CIV lines in our study are not known. We also note that optically thick lines {\it can} respond to ionization changes in inhomogeneous partial covering situations if some portions of the background light source are covered by optically thin gas \citep{Hamann11, Hamann12}. Thus the cause of the variability in our study remains uncertain. 

We consider the crossing cloud scenario in our discussions below because it yields small distances and conservatively small lower limits on the outflow masses and kinetic energies. This, in turn, leads us to conservative assessments of whether quasar outflows have sufficient energy and momentum for feedback to galaxy evolution (\S\ref{energy}). 

Following \cite{Capellupo13}, we assume the outflow crossing speeds are Keplerian at each radial distance, $R$, from the central black hole of mass, $M_{BH}$. There are, of course, other possibilities (e.g., \citealt{Hall11}, \citealt{Rogerson16}). However, without a specific model of the outflow structure, we prefer to take this simple approach. We also note that Keplerian speeds are a reasonable first guess because the outflows are launched from a rotating accretion disk. Then, as the flows expand outward, the v $\propto 1/\sqrt{R}$~ dependence of crossing speeds in Keplerian orbits is intermediate between v $\propto 1/R$~ expected from conservation of angular momentum (in a flow with only radial forces) and other schemes where strong magnetic fields threaded through the disk can add angular momentum to produce larger-than-Keplerian crossing speeds at larger radii \citep{Everett05, Fukumura10}. 

To derive specific crossing speeds and outflow locations, we need to estimate the black hole masses and the sizes of the UV emission sources. We use the luminosity to estimate a black hole mass assuming the quasars are accreting at $\sim$1/3 of the Eddington rate. We adopt a diameter for the continuum emitting region of D = 0.01 pc at 1450$\AA$ (from Hamann et al., in prep.). This estimate is based on the radial temperature structure expected for a geometrically thin, optically thick accretion disk (e.g., \citealt{Peterson97}) modified by recent results on continuum region sizes from micro-lensing studies (\citealt{Blackburne15}, \citealt{Chartas16}, and refs. therein). 

We also need to consider the absorbing geometry. \cite{Capellupo13} plot the relationship of the variability time to crossing speed and distance for two schematic geometries. We adopt their `crossing disc' model, which depicts a uniform circular absorber moving radially across a larger uniform circular emission source. The crossing speed in this schematic model is given by v$_{\rm cross} \sim \sqrt{\Delta A}\, D_{1450}/\Delta t$, where $\Delta t$ is the variability time and $\Delta A$ is the change in the normalized depth of the line trough (\S\ref{variability}) corresponding to the change in the absorber covering fraction for optically thick lines (Equation \ref{line_fit}).  

Table \ref{table:locations} lists the crossing speed and distance derived from our `typical' mini-BAL $\Delta t$ and $\Delta A$ (\S\ref{variability}) as well as some specific values of these parameters that portray the range of results in our study. For the typical mini-BAL case, we use an average bolometric luminosity of $L=3\times 10^{47}$ ergs/s (Table \ref{table:quasar_properties}) corresponding to a black hole mass of $M_{BH}\approx 7\times 10^9$ $M_{\odot}$. For the other cases, we consider the shortest timescale $\Delta t_{\mathrm{shortest}}$ and the largest amplitude $\Delta A_{\mathrm{largest}}$ change overall in the mini-BAL and BAL quasars, using the luminosities (Table \ref{table:quasar_properties}) and corresponding black hole masses for those quasars. The results in Table \ref{table:locations} are consistent with other variability studies that consider moving clouds/structures across our lines of sight \citep{Paola11, Paola13, Hall11, Capellupo13, Capellupo14, Rogerson16}. 

\subsection{Outflow Energetics}\label{energy}

Finally, we discuss the outflow energetics in the BAL quasar J1301+0002 based on the results described above. If the outflow has a radial thickness that is small compared to its radial distance from the quasar, then its total mass is given simply by
\begin{equation}\label{mass}
	M \ \ga \ 1700 \left(\frac{Q}{15~\%}\right) \left(\frac{N_H}{2.5\mathrm{x}10^{22}~\mathrm{cm^{-2}}}\right) \left(\frac{R}{2~\mathrm{pc}}\right)^2 M_{\odot}
\end{equation}
where $N_H\ga 2.5\times 10^{22}$ cm$^{-2}$ is the lower limit column density we derive from $\tau_0 \ga$ 3 in the weaker PV 1128 \AA\ line (\S\ref{structure}), $R=2$ pc is an approximate distance consistent with our `typical' variability parameters (Table \ref{table:locations}, also \citealt{Capellupo13}), and $Q=15\%$ is a nominal global covering fraction of BAL outflows (i.e. the fraction of 4$\pi$ steradians covered by the flow as seen by the central quasar) based on observed detection frequencies of BALs in quasar spectra (e.g., \citealt{Hewett03},  \citealt{Reichard03}, \citealt{Trump06},  \citealt{Knigge08}, \citealt{Gibson09a}). 

The kinetic energy in the outflow, $K = M$v$^2$/2, is then
\begin{equation}\label{ke}
	K \ \ga \ 4.3\times 10^{54} \left(\frac{M}{1700~M_{\odot}}\right) \left(\frac{\rm v}{16229~\mathrm{km}~\mathrm{s^{-1}}}\right)^2 ~ \mathrm{ergs}
\end{equation}
where v$=16299$ km/s is the velocity determined from our PV line fits. We can convert the mass and kinetic energy into a time-averaged mass loss rate, $\dot{M}$, and kinetic energy luminosity, $\langle L_k \rangle$, by dividing each by the characteristic flow time, $t_{\mathrm{flow}}\sim R/$v~$\sim 120$ yr. This yields $\dot{M}\ga 14$ M$_{\odot}$/yr, $\langle L_k \rangle \ga 1.1\times 10^{45}$ ergs/s, and a ratio of the kinetic energy luminosity to quasar bolometric luminosity of $\langle L_k \rangle$/$L \ga 0.65\%$. This lower limit is similar to other estimates of quasar outflow kinetic energies based on PV BAL detections and pc-scale distances \citep[][and in prep.]{Capellupo14}. It is also similar to estimates of the minimum outflow power needed to disrupt star formation and mass assembly in the host galaxies (e.g., $\langle L_k \rangle$/$L \ga\, 0.5$\% to 5\%, \citealt{Scannapieco04},  \citealt{DiMatteo05}, \citealt{Prochaska09}, \citealt{Hopkins10}). 

Thus it appears that quasar outflows, at least BAL outflows with PV absorption, have sufficient energies for feedback to galaxy evolution. Another way to look at this is that the total outflow kinetic energy estimated above (Eqn. \ref{ke}) is equivalent to $\ga$4300 Type II supernovae (each with energy $\sim10^{51}$ ergs). This amount of energy release is at least competitive with a major burst of star formation. We also point out that if the outflow distances are larger than the pc-scales derived from the assumption of moving clouds, as suggested by some other BAL studies (\citealt{Moe09}, \citealt{Dunn10}, \citealt{Arav13}, \citealt{Borguet13}), then our inferred lower limits on the kinetic energies would be dramatically larger (by a factor of $R^2$, Eqn. \ref{mass}). 

However, it is still not known if the large masses and kinetic energies inferred from PV measurements are common in quasar outflows. For example, we derive a typical column density of $N_H\ga 1.3\times 10^{20}$ cm$^{-2}$ using a generously large CIV $\tau_0$ value from the collection of mini-BALs and adhering to the same assumptions and analysis used for PV (see \S\ref{bal_fits}). This column density is roughly two orders of magnitude less than the BAL quasar J1301+0002 that is given in Equations \ref{mass} and \ref{ke}, but this column density could be a severe lower limit. There are some indications that PV absorption and thus large column densities are common based on four out of eight low-redshift BAL quasars showing PV in the small HST survey by \cite{Hamann03}, and the recent study by Herbst et al. (in prep.) showing that a majority of BAL and strong mini-BAL quasars have {\it saturated} PV lines (in a 1:1 doublet ratio). The Herbst et al. result suggests that mini-BALs are weaker than BALs because of smaller line-of-sight covering fractions, not lower column densities, as discussed in \S\ref{structure}. It is possible that PV absorption is present in many individual mini-BAL quasars (including our sample, \S\ref{mini_fits}) but the PV lines remain generally undetected due to problems with blending in the Ly$\alpha$ forest and the general weakness of PV, e.g., compared to the already weak CIV in mini-BAL systems.  

\section{Summary} \label{summary}
We acquired HST and multi-epoch ground-based spectra for a sample of seven quasars that have a range of CIV outflow lines from narrow/weak mini-BALs to a single quasar with a broad/strong BAL. The wavelengths covered by our study are from at least 850 $\AA$ to 1650 $\AA$ in the quasar rest frame. This is the first study to target specifically mini-BALs and weak BALs in the far-UV for comparisons to the stronger and more often measured BALs. We use these data to study the outflow ionizations, column densities, locations, and kinetic energies. Our main results are the following.

\begin{enumerate}
	\item \textit{Lines Detected:} Broad OVI outflow lines are detected in every quasar. Our study builds upon \cite{Paola11} by, for the first time, not only adding detections of OVI in mini-BALs systems, but also finding for the first time that OVI is generally stronger than CIV. In the six mini-BAL quasars, the OVI REWs are 2--6 times stronger than CIV indicating that high degrees of ionization are generally favored. The BAL system in J1301+0002 includes a wider variety of lines including PV, SVI, and SiIV. Low-ionization outflow lines such as CII are not detected in any quasar. 
		
	\item \textit{Outflow Structure:} The OVI lines in at least three of the six mini-BAL quasars are optically thick based on $\sim$1:1 doublet ratios. The lines do not reach zero intensity indicating that the absorber only partially covers the background light source along our lines of sight (\S\ref{mini_fits}, \S\ref{structure}). In the BAL quasar, all of the measured BALs are optically thick and their different depths in the observed spectrum correspond to covering fractions from $\sim$0.27 to $\sim$0.75 in a complex inhomogeneous outflow (\S\ref{bal_fits}, \S\ref{structure}). The partial covering requires absorbing region sizes smaller than the UV continuum source $\la$ 0.01 pc across. The highest-column density gas represented by PV absorption in the BAL quasar is the most spatially compact across the most limited velocity range (leading to weaker and narrower line profiles compared to the other BALs). We speculate that the mini-BALs and BALs in our study form in similar clumpy or filamentary outflow structures, and that the mini-BALs identify smaller or fewer outflow clumps along the lines of sight (\S\ref{structure}). 

	\item \textit{CIV Variability:} The broad CIV outflow lines varied in all seven quasars within a time frame of $\leq$ 1.9 yrs in the rest frame (\S\ref{variability}). We estimate that a `typical' mini-BAL variability time is $\Delta t \sim 1.1$ yrs (based on half of the mini-BAL quasars varying in that time) with a typical amplitude change in the normalized line depths of $\Delta A\sim 0.12$. These results are consistent with previous variability studies of BALs and mini-BALs (\S\ref{variability}).
	
	\item \textit{An Emerging High-Velocity BAL:} We discovered the emergence of a high velocity BAL at $-$38,800 km/s that accompanied the strengthening of a previously known mini-BAL at $-$22,800 km/s in the quasar J1031+3807 (\S\ref{variability}). Cases like this of coordinated dramatic line variabilities support the idea that BALs and mini-BALs are closely related manifestations of the same general outflow phenomenon (\S\ref{structure}).
	
	\item  \textit{Outflow Locations:} We estimate outflow locations assuming the line variations are caused by clumps or filaments crossing our lines of sight at speeds similar to the Keplerian velocity (\S\ref{locations}). `Typical' mini-BAL variability parameters thus yield a typical radial distance of $R\sim 2$ pc. However, we cannot rule out the possibility that the line variations are caused by changes in the outflow ionization, which leads to very large upper limits on the outflow distances of order $\sim$1 kpc. 
	
	\item \textit{Outflow Mass and Kinetic Energy:} Our estimate of minimum PV optical depth $\tau_0(1128) \ga$ 3 in the BAL quasar J1301+0002 (based on line fits and an observed $\sim$1:1 doublet ratio, \S\ref{bal_fits}) corresponds to a total column density of $2.5\times 10^{22}$ cm$^{-2}$ in standard photoionization models with solar abundances (\S\ref{structure}). Combining this lower limit with our estimated outflow location $R\sim 2$ pc (\S\ref{locations}) indicates that the total outflow mass and kinetic energy are $\ga$14 M$_{\odot}$/yr and $\ga$$4.3\times 10^{54}$ ergs/s, respectively (\S\ref{energy}). Dividing by a characteristic flow time yields a time-averaged kinetic energy luminosity compared to bolometric, $\left<L_K\right>/L \ga 0.65$\%, which is in the range believed to be sufficient for feedback to contribute to galaxy evolution. 
\end{enumerate}

\section*{Acknowledgements}
We are grateful to Martin Durant for providing software to extract the HST COS spectra. This work was supported in part by a grant from the Space Telescope Science Institute GO-11705. We thank an anonymous referee for helpful comments on the manuscript.







\bibliographystyle{mnras}
\bibliography{bibliography}
\clearpage




\bsp	
\label{lastpage}
\end{document}